%% file: main.tex
\documentclass[aps,prd,reprint,nofootinbib,superscriptaddress,flushbottom]{revtex4-1}

\usepackage{newtxtext}
\usepackage[varg]{newtxmath}
\usepackage[T1]{fontenc}
\usepackage{graphicx}

\InputIfFileExists{stylesheet.tex}{}{}

\usepackage[
	hyperfootnotes=false,
	colorlinks=true,
	allcolors=blue,
	breaklinks=true]
{hyperref}
\newcommand{\dx}{\text{d}}
\DeclareMathOperator{\sgn}{sgn}

\begin{document}
\title{Constraining spontaneous black hole scalarization in\protect\\scalar-tensor-Gauss-Bonnet theories with current gravitational-wave data}

\author{Leong Khim \surname{Wong}}
\affiliation{Universit\'{e} Paris-Saclay, CNRS, CEA, Institut de Physique Th\'{e}orique, 91191 Gif-sur-Yvette, France}

\author{Carlos A. R. \surname{Herdeiro}}
\author{Eugen \surname{Radu}}
\affiliation{Departamento de Matem\'{a}tica da Universidade de Aveiro and
Centre for Research and Development in Mathematics and Applications (CIDMA),
Campus de Santiago, 3810-183 Aveiro, Portugal}

\begin{abstract}
We examine the constraining power of current gravitational-wave data on scalar-tensor-Gauss-Bonnet theories that allow for the spontaneous scalarization of black holes. In the fiducial model that we consider, a slowly rotating black hole must scalarize if its size is comparable to the new length scale $\lambda$ that the theory introduces, although rapidly rotating black holes of any mass are effectively indistinguishable from their counterparts in general relativity. With this in mind, we use the gravitational-wave event GW190814\,---\,whose primary black hole has a spin that is bounded to be small, and whose signal shows no evidence of a scalarized primary\,---\,to rule out a narrow region of the parameter space. In particular, we find that values of ${\lambda \in [56,96]~M_\odot}$ are strongly disfavored with a Bayes factor of~$0.1$ or less. We also include a second event, GW151226, in our analysis to illustrate what information can be extracted when the spins of both components are poorly measured.
\end{abstract}

\maketitle

\section{Introduction}
\label{sec:intro}

General relativity~(GR) is one of the crowning achievements of modern physics.~Yet, the theory is famously ultraviolet incomplete and, moreover, is in itself unable to elucidate the precise nature of some of the key constituents of our Universe, like dark matter and dark energy. These shortcomings have inspired numerous proposals for gravitational theories beyond~GR, with some being better constrained by observational data than others. (See Refs.~\cite{Will:2014kxa, Berti:2015itd, Yunes:2016jcc, Damour:PDGReview} for reviews.)

From a bottom-up perspective, we know that the microscopic details of any potential ultraviolet completion must manifest as higher-dimensional operators in the low-energy effective action. Adding higher-curvature terms to the Einstein-Hilbert action, however, often introduces theoretical issues, like runaway modes or ghosts, on account of the field equations becoming greater than second~order~\cite{Ostrogradsky:1850fid}. One way to evade these issues is to focus only on certain curvature combinations known as Euler densities, which for purely gravitational theories would leave us with the particular case of Lovelock gravity~\cite{Lovelock:1971yv}. The first such combination (beyond the cosmological constant and the Ricci scalar) is the Gauss-Bonnet (GB) quadratic curvature invariant. Including this term in the effective action is also well motivated from the context of string theory, wherein the GB invariant appears naturally as a correction to Einstein's theory~\cite{Zwiebach:1985uq}. A~bare GB term is dynamical only in five or more spacetime dimensions (${D>4}$), however, and thus a common tactic to render it dynamical also in ${D=4}$ is to couple it to a new scalar degree of freedom~$\phi$, to which we confer a canonical kinetic term. The result is known as a scalar-tensor-Gauss-Bonnet~(STGB)~theory.

Over the years, black hole~(BH) solutions in a variety of STGB theories have been studied in the literature. Arguably the simplest of these theories are the dilatonic~\cite{Kanti:1995vq, Kleihaus:2011tg} and shift-symmetric Horndeski models~\cite{Sotiriou:2014pfa, Delgado:2020rev}, although notably neither admit the familiar Schwarzschild or Kerr geometries of vacuum GR as elementary BH solutions. Instead, BHs in these theories are described by new geometries that sport a nontrivial scalar-field profile outside the horizon. Often, these solutions are referred to as BHs with scalar ``hair''~\cite{Herdeiro:2015waa}.

More recently, it was realized that there exists another family of STGB theories that admits both vacuum-GR BHs \emph{and} new hairy BHs as valid solutions. A subset of this family, moreover, makes the former type of BH unstable to scalar perturbations (in certain regions of parameter space) and thereby provides a mechanism by which the latter can form dynamically. This is the phenomenon of ``spontaneous scalarization'' in STGB theories~\cite{Doneva:2017bvd, Silva:2017uqg, Antoniou:2017acq}, which is itself inspired by a similar process, first discussed in the 1990s, that occurs with neutron stars in simpler scalar-tensor~models~\cite{Damour:1993hw}.

On the observational side, what is most interesting about spontaneous scalarization is that it offers a distinctive signature of new physics in the strong-gravity regime, but only for \emph{some}~BHs, specifically, those whose Schwarzschild radii are comparable to the new length scale~$\lambda$ that the theory introduces. Heavier and lighter BHs, meanwhile, remain indistinguishable from their counterparts in~GR. Although this does mean that these theories are inherently quite challenging to constrain (arguably more so than, say, dilatonic or shift-symmetric STGB theories~\cite{Wang:2021yll, Perkins:2021mhb, Lyu:2022gdr}), our aim in this paper is to show that this is, nevertheless, already possible with current gravitational-wave~(GW)~data. We substantiate this claim by focusing on a fiducial STGB theory and confronting it with two events: GW190814~\cite{LIGOScientific:2020zkf} and GW151226~\cite{LIGOScientific:2016sjg}. Because a larger spin quenches the extent to which a spontaneously scalarized BH deviates from Kerr~\cite{Cunha:2019dwb}, we shall see that the former event\,---\,whose primary BH has a spin that is bounded to be small\,---\,is able to provide a robust constraint on the new length scale~$\lambda$. As a point of comparison, we include GW151226 in our analysis to illustrate what information can be extracted from GWs when the spins of both components in the binary are~poorly~measured.

\begin{figure*}
\includegraphics{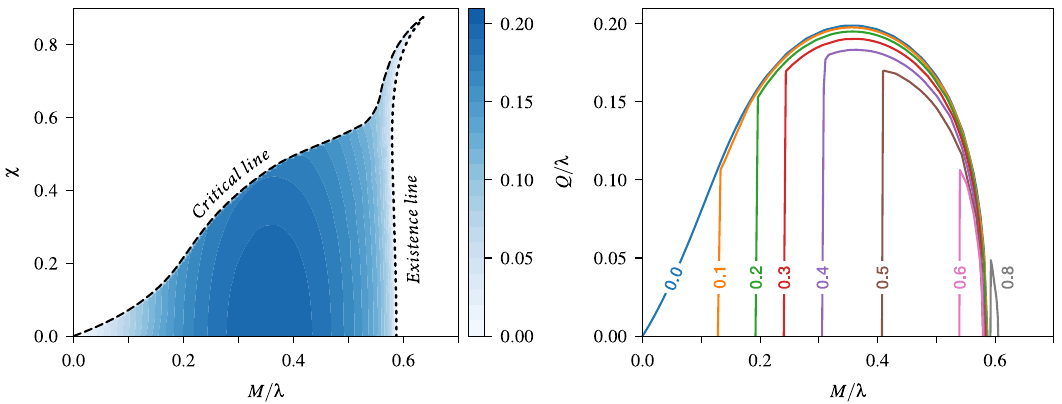}
\caption{Family of scalarized black hole solutions in the scalar-tensor-Gauss-Bonnet theory of Eqs.~\eqref{eq:stgb_action}~and~\eqref{eq:stgb_coupling_function}. Left: the dimensionless scalar charge $Q/\lambda$ as a function of the dimensionless mass~$M/\lambda$ and dimensionless spin~$\chi$. Right: $Q/\lambda$ as a function of $M/\lambda$ for different values of~$\chi$ between $0.0$ and $0.8$.}
\label{fig:scalar_charge}
\end{figure*}

The remainder of this paper proceeds as follows. We begin in Sec.~\ref{sec:stgb} by providing a brief overview of the specific STGB theory under consideration and the key properties of its scalarized~BHs. In Sec.~\ref{sec:gw}, we then describe our Bayesian approach for establishing constraints using GW data. Our likelihood and waveform model are presented in Sec.~\ref{sec:gw_likelihood}, while our choice of prior and some additional comments about the GW events being used are discussed in Sec.~\ref{sec:gw_prior}. The main results of this study can be found in Sec.~\ref{sec:gw_results}, with some ancillary details relegated to an~Appendix. We conclude in Sec.~\ref{sec:discussion}.

\section{Scalar-tensor-Gauss-Bonnet theory}
\label{sec:stgb}

When written with units in which ${G=c=1}$, the action for any STGB theory reads
\begin{equation}
	S = \frac{1}{16\pi}\int\dx^4x\sqrt{-g}
	\big[
		R - 2\partial_\mu\phi \partial^\mu\phi + \lambda^2 f(\phi)\mathcal{G}
	\big].
\label{eq:stgb_action}
\end{equation}
The quantity ${\mathcal G \equiv R^{\mu\nu\rho\sigma} R_{\mu\nu\rho\sigma} -4R^{\mu\nu} R_{\mu\nu} + R^2}$ is the GB invariant, $\lambda$ is a coupling constant with dimensions of length, and $f(\phi)$ is an arbitrary function of the real scalar field~$\phi$, which we are free to specify.

A dilatonic STGB theory corresponds to choosing ${f(\phi) = e^{\alpha \phi} }$, where ${\alpha \neq 0}$ is a dimensionless constant, whereas the choice ${ f(\phi)=\alpha \phi}$ gives us a shift-symmetric theory. Common to these two choices is the fact that the first derivative of the coupling function $f'(\phi)$ is nonvanishing for any (finite) value of~$\phi$; hence,
the scalar field equation
${ \nabla^\mu\nabla_\mu \phi = - \lambda^2 f'(\phi)\mathcal{G}/4 }$
cannot be solved by having $\phi$ equal to a constant whenever ${\mathcal G \neq 0}$.~This implies that the Schwarzschild and Kerr geometries of vacuum GR are not solutions to this class of theories, as previously discussed in~the~Introduction.

To obtain a theory that allows for the spontaneous scalarization of BHs, we want a coupling function whose first two derivatives satisfy ${f'(\phi_0) = 0}$ and ${f''(\phi_0) \neq 0}$ for some value~$\phi_0$~\cite{Doneva:2017bvd, Silva:2017uqg}, which we can take to be zero without loss of generality. When these two conditions are met, a Kerr spacetime with ${\phi = 0}$ is always a solution to the field equations, but scalarized BHs can coexist in certain regions of the parameter space, namely, wherever the Kerr metric is linearly unstable to scalar perturbations. The ensuing tachyonic instability triggers a strong-gravity phase transition that promotes the growth of a scalar-field profile around the horizon and results in a new, scalarized equilibrium~solution that, for appropriate choices of the coupling function, is entropically~favored.

Two common choices for $f(\phi)$ that satisfy these scalarization requirements are the quadratic monomial ${f(\phi) = \beta\phi^2}$~\cite{Silva:2017uqg} (see~also~Refs.~\cite{Blazquez-Salcedo:2018jnn, Silva:2018qhn, Macedo:2019sem, Antoniou:2021zoy, Herrero-Valea:2021dry}) and the Gaussian-like function~\cite{Doneva:2017bvd, Cunha:2019dwb}
\begin{equation}
	f(\phi) = \frac{1}{2\beta}(1-e^{-\beta\phi^2}),
\label{eq:stgb_coupling_function}
\end{equation}
where $\beta$ is a dimensionless parameter. For illustrative purposes, we shall focus only on the latter option in this paper and, following Refs.~\cite{Doneva:2017bvd, Cunha:2019dwb}, will further fix ${\beta = 6}$ for simplicity.

The general properties of scalarized BHs in this theory were previously explored in Ref.~\cite{Cunha:2019dwb} by interpolating between a discrete set of about a thousand numerical solutions. Because the field equations are invariant under a global rescaling of ${\lambda \to b\lambda}$ and ${r \to b r}$ (where~$r$ is the radial coordinate and ${b>0}$ is an arbitrary constant), this family of BH solutions can be parametrized by the single length scale~$\lambda$ and two dimensionless~parameters:
\begin{equation}
	M/\lambda
	\quad\text{and}\quad
	\chi \equiv  S/M^2.
\end{equation}
The symbol $M$ denotes the (Arnowitt-Deser-Misner) mass of the BH, while $S$ is the magnitude of its spin~angular~momentum.

The left panel of Fig.~\ref{fig:scalar_charge} shows the domain of existence for these scalarized BHs, which is bounded by three curves in the $M/\lambda$--$\chi$ plane: an existence line, a critical line, and the line of static, scalarized solutions (which has~${\chi=0}$). The existence line coincides with the set of bifurcation points along which these scalarized BHs branch off from the Kerr family, whereas the critical line marks the boundary of this space of solutions beyond which the repulsive effect from the GB term prohibits the existence of an event horizon~\cite{Cunha:2019dwb}.

A useful quantity that characterizes how hairy these scalarized solutions are, and therefore how different they are from Kerr, is the scalar ``charge''~$Q$, which is read off from the far-field, asymptotic profile of the scalar field as~${\phi \sim -Q/r}$. The aforementioned scaling symmetry implies that $Q/\lambda$ is a dimensionless function of the two parameters $M/\lambda$ and~$\chi$. This dependence is shown as a heat map in the left panel of Fig.~\ref{fig:scalar_charge}, while the right panel provides an alternative visualization by plotting $Q/\lambda$ against $M/\lambda$ for different fixed values~of~$\chi$. (As~this theory admits a ${\phi \to -\phi}$ symmetry, it suffices to discuss only those solutions for which ${Q > 0}$.)

Naturally, the exact way in which $Q/\lambda$ varies with $M/\lambda$ and $\chi$ depends on our choice of~$f(\phi)$, as does the precise shape and location of the critical line.~The existence line, however, is \emph{universal} to any coupling function that satisfies ${f'(0) = 0}$ and ${f''(0) > 0}$,\footnote{In this case, we can always redefine $\lambda$ such that ${f''(0) = 1}$.} as a linear stability analysis of scalar perturbations about the Kerr metric is all that is required to determine its location~\cite{Doneva:2017bvd, Silva:2017uqg, Herdeiro:2018wub,Andreou:2019ikc, Astefanesei:2020qxk, Ventagli:2020rnx, Hod:2019pmb, Herdeiro:2019yjy, Annulli:2022ivr}. For slowly rotating BHs, one finds that the threshold value of $M/\lambda$ below which spontaneous scalarization can occur~is\looseness=-1
\begin{equation}
	M/\lambda \simeq 0.587,
\label{eq:stgb_threshold}
\end{equation}
although this value increases somewhat for larger spins. It is worth mentioning, in fact, that the existence and critical lines actually extend all the way up to spins close to the Kerr bound,\footnote{We know this because $\mathcal{G}_\text{Kerr}$ remains positive in the vicinity of the equator even when ${\chi \sim 1}$~\cite{Cherubini:2002gen}, and so can continue to trigger the tachyonic instability responsible for spontaneous scalarization.} contrary to what Fig.~\ref{fig:scalar_charge} might suggest. The region sandwiched between these two lines becomes increasingly narrow, however, and is therefore difficult to resolve numerically. For our purposes, this is not a problem, as this tiny region would not be resolvable by the inherently noisy GW data anyway; hence, we can effectively assume that any BH with ${\chi \gtrsim 0.88}$ or ${M/\lambda \gtrsim 0.64}$ is~always~unscalarized.

To summarize, the main takeaway of this section is that a BH will spontaneously scalarize only if its size is comparable to the coupling constant~$\lambda$ in the theory. For a BH with a given mass~$M$, the specific range of values of~$\lambda$ for which this can occur strongly depends on the magnitude of the spin~$\chi$, with smaller spins resulting in wider ranges. Since these scalarized BHs are entropically favored over their counterparts in~GR, any BH that is observed to be indistinguishable from Kerr can be used to rule out the values of~$\lambda$ for which scalarization should have occurred. This is the basic argument that we will use in the next section to~establish~constraints.

\section{Gravitational-wave constraints}
\label{sec:gw}

A binary system with one or more scalarized components necessarily radiates energy into both gravitational and scalar waves. By working under the hypothesis that GR accurately describes all known GW events to date~\cite{TheLIGOScientific:2016src, LIGOScientific:2018dkp, LIGOScientific:2019fpa, LIGOScientific:2020tif, LIGOScientific:2021sio}, we can translate the absence of any measurable scalar radiation into a constraint on the value of the coupling constant~$\lambda$. This information is encoded in the posterior density~$p(\lambda|d)$ for~$\lambda$ given some data~$d$, which we shall obtain in this section by first computing the full posterior~$p(\lambda,\vartheta|d)$ and then marginalizing over all of the other parameters~$\vartheta$ in the waveform model. Writing ${\theta \equiv (\lambda,\vartheta)}$ for brevity, Bayes' theorem tells us that
\begin{equation}
	p(\theta | d)
	\propto
	p(d|\theta) \pi(\theta),
\end{equation}
where $p(d|\theta)$ is the likelihood of the data given the model para\-meters, while $\pi(\theta)$ is the prior for these~model~parameters.

\subsection{Likelihoods and waveform models}
\label{sec:gw_likelihood}

It is common practice to assume that the noise in each detector is stationary, Gaussian, and uncorrelated~\cite{LIGOScientific:2019hgc} such that the likelihood for a single event is~\cite{Cutler:1994ys, Veitch:2014wba}
\begin{equation}
	p(d|\theta) \propto\exp
	\bigg(
		{-2} \sum_{i\mathstrut}\int_{f_{\,\text{low},i}}^{f_{\,\text{high},i}}\!
		\frac{|\tilde d_i(f) - \tilde h_i(f;\theta)|^2}{S_{n,i}(f)}
		\dx f
	\bigg).
\label{eq:gw_likelihood}
\end{equation}
For the $i$th detector, $\tilde d_i(f)$ and $\tilde h_i(f;\theta)$ represent the Fourier transforms of the signal and the associated waveform model projected onto the detector, respectively, while $S_{n,i}(f)$ denotes the power spectral density of the~noise.

Given our working hypothesis, waveform models for STGB theories may easily be constructed from their GR counterparts via the inclusion of small correction terms that account for the would-be effects of the scalar field~\cite{Yunes:2009ke}. Said in other words, if $\tilde h_{\ell m}^\text{(GR)}(f;\vartheta)$ is a (spin-weighted) spherical harmonic mode of the GR waveform,~then
\begin{equation}
	\tilde h_{\ell m}^{\mathstrut}(f;\theta)
	=
	\tilde h^\text{(GR)}_{\ell m}(f;\vartheta)\, e^{-i\delta\phi_{\ell m}(f;\theta)}
\label{eq:gw_waveform_deformations}
\end{equation}
is a good approximation to the corresponding mode in STGB gravity.~The form of~$\delta\phi_{\ell m}$ during the inspiral stage of the coalescence is well known from numerous studies of general scalar-tensor theories~\cite{Damour:1992we, Mirshekari:2013vb, Lang:2013fna, Lang:2014osa, Bernard:2018hta, Bernard:2018ivi, Sennett:2016klh, Shao:2017gwu, Huang:2018pbu, Kuntz:2019zef, Brax:2021qqo, Bernard:2022noq, Yagi:2011xp, Julie:2019sab, Shiralilou:2020gah, Shiralilou:2021mfl}, and is given by~\cite{Shao:2017gwu}
\begin{equation}
	\delta\phi_{\ell m}(f;\theta)
	=
	\frac{5m }{14\,336\,\eta}
	\bigg(
		\frac{Q_1}{M_1}-\frac{Q_2}{M_2}
	\bigg)^{\!\!2}
	\bigg( \frac{2\pi Mf}{m} \bigg)^{\!\!-7/3}
\label{eq:gw_waveform_correction_term}
\end{equation}
at leading order in the post-Newtonian~(PN) expansion. In~the above, $Q_1$ and $Q_2$ are the scalar charges of the two components, ${M=M_1 + M_2}$ is the total mass of the binary, and ${\eta = M_1 M_2/M^2}$ is its symmetric mass ratio.

The higher-order terms in $\delta\phi_{\ell m}$ are presently known up to relative $+2$PN~order~\cite{Sennett:2016klh}, but we expect that using just the leading ($-1$PN-order) term in Eq.~\eqref{eq:gw_waveform_correction_term} will suffice to establish a conservative bound on~$\lambda$. Indeed, this expectation is supported by recent Bayesian analyses of dilatonic and shift-symmetric STGB theories~\cite{Lyu:2022gdr, Perkins:2022fhr}, which found that the bounds on~$\lambda$ improve only marginally when higher-order terms are included. It is for this same reason that we have elected to focus only on the corrections to the phase of the waveform in Eq.~\eqref{eq:gw_waveform_deformations}; corrections to the amplitude may be neglected as a first approximation~\cite{Tahura:2019dgr}. Having said all of this, we will still make use of one important piece of insight gleaned from the study of these higher-order terms: while BHs in isolation can spontaneously scalarize to produce a charge~$Q$ of either sign, it has been shown that binary systems with oppositely charged components are unstable and cannot inspiral towards merger adiabatically, if at all~\cite{Sennett:2017lcx, Julie:2022huo}. Since the signals to be analyzed are known to be products of binary systems that have undergone a successful merger, we shall restrict ourselves to waveforms with ${\sgn(Q_1) = \sgn(Q_2)}$~in~what~follows.

Unlike the inspiral, much less is known about the merger and ringdown stages of a binary in STGB gravity, or any scalar-tensor theory, for that matter (but see Refs.~\cite{Carson:2020ter, Witek:2018dmd, Witek:2020uzz, Silva:2020omi, East:2020hgw, East:2021bqk, Doneva:2022byd, Okounkova:2017yby, Okounkova:2019dfo, Okounkova:2019zjf, Okounkova:2020rqw, Bonilla:2022dyt} for recent progress). Nonetheless, we can proceed in the absence of such information by setting the high-frequency cutoff $f_\text{high}$ in Eq.~\eqref{eq:gw_likelihood} equal to the maximum frequency of the inspiral stage [defined in Eq.~(5.8) of Ref.~\cite{Pratten:2020fqn}], thereby excluding the merger and ringdown portions of the signal from our analysis. Discarding information in this manner naturally comes at the cost of precision, but because the effects of scalar radiation are more pronounced during the early stages of the inspiral [observe that Eq.~\eqref{eq:gw_waveform_correction_term} is larger at lower frequencies], we do not expect our constraints to change dramatically when a full inspiral-merger-ringdown analysis becomes possible. As for the low-frequency cutoff, we will generally set ${f_\text{low} = 20~\text{Hz}}$, except when data quality considerations require a different value (see~Sec.~\ref{sec:gw_prior}~below).

It remains to discuss how one obtains the projected waveform $\tilde h_i(f;\theta)$ from the harmonic modes in Eq.~\eqref{eq:gw_waveform_deformations}. We begin by using the \mbox{IMRPhenomXHM} model of Ref.~\cite{Garcia-Quiros:2020qpx} to generate $\tilde h^\text{(GR)}_{\ell m}(f;\vartheta)$ in the so-called ``$L$-frame'' of the binary, wherein the orbital angular momentum~$\vec{L}$ points along the positive $z$~axis. Note, however, that this frame is generically noninertial due to the precession of~$\vec{L}$ about the total angular momentum~$\vec{J}$. After multiplying by~$e^{-i\delta\phi_{\ell m}}$, we use the ``twisting-up'' procedure implemented in the \mbox{IMRPhenomXPHM} model~\cite{Pratten:2020ceb} to map $\tilde h_{\ell m}(f;\theta)$ onto the ``plus'' and ``cross'' polarizations of the waveform in the inertial $J$-frame. (The twisting-up procedure that is appropriate to STGB gravity will in principle differ from that of GR, but these differences are minor%
\footnote{The precession of $\vec{L}$ around $\vec{J}$ is determined by solving appropriate PN equations of motion~\cite{Pratten:2020ceb}. In scalar-tensor theories, the corrections to these equations due to the scalar field are proportional to $(Q_1 / M_1) (Q_2 / M_2)$~\cite{Brax:2021qqo}, and so are small under our working hypothesis that GR provides an accurate description of~the~data.}
and, as we did with the higher-order terms in~$\delta\phi_{\ell m}$, can be neglected as a first approximation.) The remaining projection of the polarization modes onto the detector is straightforward~\cite{Cutler:1994ys} and is handled automatically for us by the Bilby~software~package~\cite{Ashton:2018jfp, Romero-Shaw:2020owr}.

\subsection{Events and priors}
\label{sec:gw_prior}

In Sec.~\ref{sec:stgb}, we established that the range of values of $\lambda$ within which a BH of mass~$M$ can spontaneously scalarize strongly depends on the magnitude of its spin, with smaller spins resulting in wider ranges. It therefore stands to reason that the strongest constraints on~$\lambda$ will come from those events in which at least one of the two spins is bounded to be small. To illustrate this point, we concentrate in this paper on the analysis of two events: GW190814~and~GW151226.

The former is one of only several events thus far with a primary BH whose dimensionless spin magnitude~$\chi_1$ is known to be small (the 90\% upper limit is ${\chi_1 < 0.07}$)~\cite{LIGOScientific:2020zkf}. Consequently, our \emph{a~priori} expectation is that the likelihood for this event will strongly disfavor a range of values of $\lambda$ near the threshold value ${M_1 / 0.587 \sim 39.5~M_\odot}$, above which the primary BH is expected to scalarize [recall~Eq.~\eqref{eq:stgb_threshold}]. The other event, GW151226~\cite{LIGOScientific:2016sjg}, is included in our analysis to serve as a point of comparison: we do not expect a strong constraint on~$\lambda$ in this case, given that both of the component spins are poorly measured. Of course, many GW events have poorly measured spins, and so the other factor that influenced our choice of GW151226 is its relatively low chirp mass. A~low chirp mass implies that the inspiral portion of the signal will last for long enough that parameter estimation can still be performed reliably (albeit with larger uncertainties) in the absence of a model for the merger and ringdown.

Data segments lasting $16\,\text{s}$ for GW190814 and $8\,\text{s}$ for GW151226 were pulled from the GW Open Science~Center database~\cite{LIGOScientific:2019lzm} and Fourier transformed to yield the signals $\tilde d_i(f)$, while longer data segments lasting $512\,\text{s}$ around the time of each event were used to determine the noise power spectral densities~$S_{n,i}(f)$~\cite{Chatziioannou:2019zvs}. As previously discussed, we fix the low-frequency cutoff in Eq.~\eqref{eq:gw_likelihood} at~$20~\text{Hz}$, except for the LIGO~Livingston data around the time of~GW190814, for which we set ${f_\text{low} = 30~\text{Hz}}$ in order to excise low-frequency acoustic noise induced by~thunderstorms~\cite{LIGOScientific:2020zkf}.

For each event, the priors for the binary's component masses and spins, along with its time and phase of coalescence, sky localization, inclination, polarization angle, and luminosity distance (i.e., all of the parameters that we have been denoting collectively by~$\vartheta$) are taken to be the same as that of Ref.~\cite{gwtc1}, while the prior distribution for the length scale~$\lambda$ is taken to be log~uniform with compact support in the range ${[1, 10^3]~M_\odot}$. Although smaller and larger values of $\lambda$ are by no means precluded, we have restricted our analysis to this finite range\,---\,in which we expect most of the interesting features of the posterior to reside\,---\,for the sake of reducing computational~cost.

Two final comments for this subsection are in order. First, note that all length scales appearing in the waveform model of Eq.~\eqref{eq:gw_waveform_deformations} are redshifted owing to the expansion of the Universe. To recover the intrinsic values of $M_1$, $M_2$, and $\lambda$ in the rest frame of the binary, we multiply each of these parameters by a factor of~$1/(1+z)$. The  redshift~$z$ is inferred from the luminosity distance by assuming a flat $\Lambda$CDM cosmology with Hubble constant and matter density parameter given by Ref.~\cite{Planck:2015fie}.

Second, while the primary component of GW190814 is massive enough to be securely identified~as a BH, its secondary\,---\,whose mass registers at around~${2.59~M_\odot}$~\cite{LIGOScientific:2020zkf}\,---\,is consistent with being either a BH or a neutron star. Admittedly, ${2.59~M_\odot}$ is rather on the high side for a neutron star, and the authors of Ref.~\cite{LIGOScientific:2020zkf} used this as a basis to argue, given what is currently known about the neutron-star equation of state, that this event is more likely to be the result of a binary BH coalescence. In this paper, we entertain only this likelier option (that GW190814's secondary is a BH) for the sake of simplicity, although it would certainly be interesting to examine the neutron star case in future work. (For~GW151226, both components are massive enough that they are almost surely BHs.)
\looseness=-1

\begin{figure}
\includegraphics{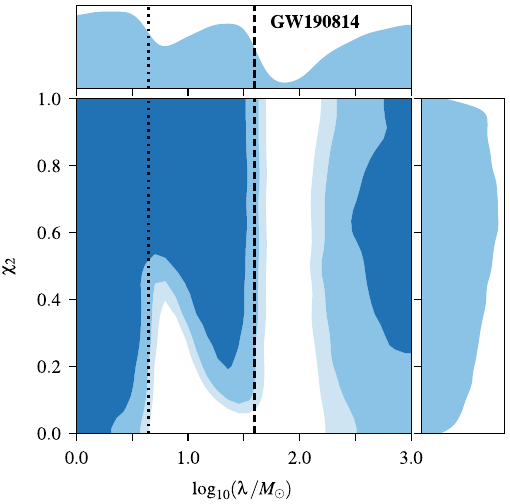}
\caption{Marginalized posterior distributions for the logarithm of~$\lambda$ and the secondary spin~$\chi_2$ in GW190814. From darkest to lightest, the shaded contours in the two-dimensional plot successively enclose 68\%, 90\%, and 95\% of the total probability in the depicted region. The dashed (dotted) vertical line marks the threshold value of $\lambda$ above which the primary (secondary) black hole is expected~to~scalarize.}
\label{fig:posterior_GW190814}
\end{figure}

\subsection{Results}
\label{sec:gw_results}

We sampled points from the posterior distribution~$p(\theta|d)$ by utilizing the Bilby-specific implementation of Dynesty~\cite{dynesty}, which is a software package for performing nested sampling \cite{Skilling:2004nsp, Skilling:2006nsp, Higson:2019dns}. For each event, we ran eight independent sampling runs\,---\,each initialized with 2000 live points\,---\,that were then merged together to produce the final output.\looseness=-1

\begin{figure*}
\includegraphics{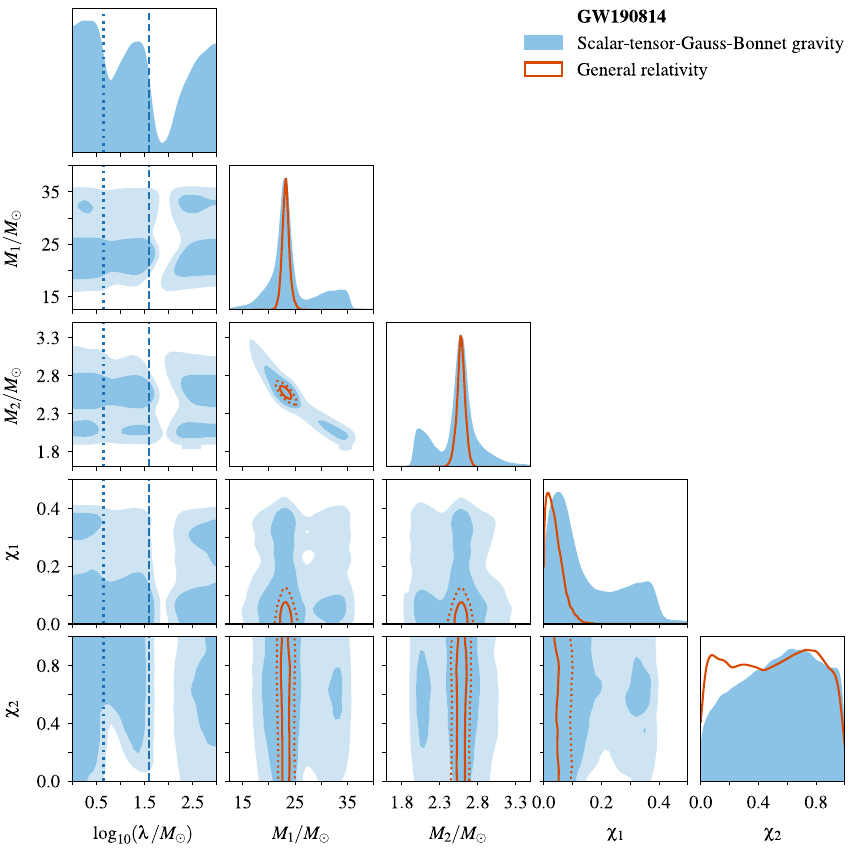}
\caption{Marginalized posterior distributions for several of the parameters used to model~GW190814. The dark and light blue contours in the two-dimensional plots enclose, respectively, 68\% and 95\% of the posterior for the scalar-tensor-Gauss-Bonnet model, while the solid and dotted orange lines enclose 68\% and 95\% of the posterior assuming general relativity (in which case,~${\lambda=0}$). In the one-dimensional plots, the posteriors for the two models have been rescaled to have the same maximum height. The blue dashed (dotted) vertical lines in the first column mark the threshold value of~$\lambda$ above which the primary (secondary) black hole is expected to scalarize.}
\label{fig:posterior_GW190814_extended}
\end{figure*}

The posterior probability for~$\lambda$ given GW190814 is shown in Fig.~\ref{fig:posterior_GW190814}, with supplementary plots presented in Fig.~\ref{fig:posterior_GW190814_extended} of the~Appendix. In both of these diagrams, the vertical dashed line at ${M_1 / 0.587 \sim 39.5~M_\odot}$ marks the threshold value of~$\lambda$ above which the primary BH is expected to scalarize, while the dotted line at ${ M_2 / 0.587 \sim 4.4~M_\odot}$ marks the corresponding value of~$\lambda$ (assuming a sufficiently small spin) for the secondary BH. Now reading Fig.~\ref{fig:posterior_GW190814} from left to right, we see that the posterior density naturally starts off at a high value because both BHs are unscalarized when~${\lambda < 4.4~M_\odot}$, and are thus identical to their counterparts in GR. For values of~$\lambda$ in between the two vertical lines, $p(\log\lambda|d)$ dips only slightly because the extent to which the secondary is scalarized strongly depends on the value~of~$\chi_2$, which is left mostly unconstrained by the data. In contrast, because $\chi_1$ is very well constrained by the data (see Fig.~\ref{fig:posterior_GW190814_extended}), the posterior strongly disfavors a range of values of~$\lambda$ above $39.5~M_\odot$ on account of there being no measurable evidence for a scalarized primary. Finally, observe that $p(\log\lambda|d)$ begins to rise again for values of ${\lambda \gtrsim 200~M_\odot}$ due to the fact that the scalar charge~$Q$ decreases with increasing~$\lambda$ once it passes its maximum~point~(see~Fig.~\ref{fig:scalar_charge}).

\begin{figure}
\includegraphics{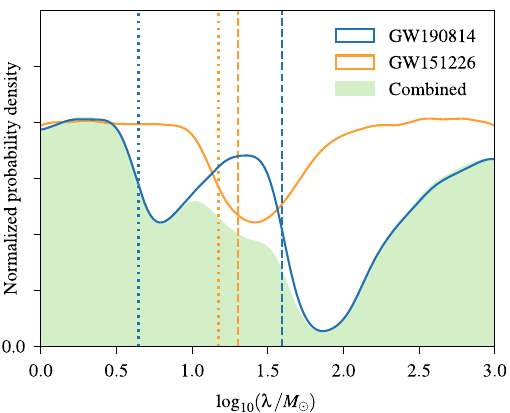}
\caption{Marginalized posterior distributions for the logarithm of~$\lambda$ as determined by GW190814 and GW151226. The details of the normalization are described in the main text. As in Fig.~\ref*{fig:posterior_GW190814}, the dashed (dotted) vertical lines mark the threshold values of~$\lambda$ above which the appropriate primary (secondary) black hole is expected to scalarize.\looseness=-1}
\label{fig:posterior_combined}
\end{figure}

In Fig.~\ref{fig:posterior_combined}, we show the marginalized posterior densities for~$\lambda$ as determined by independent analyses of GW190814 (the blue curve) and GW151226 (the orange curve) alongside the combined result from a joint analysis of both events (shaded in green). The blue vertical lines mark the same threshold values of~$\lambda$ as in Fig.~\ref{fig:posterior_GW190814}, while the orange lines at $14.9~M_\odot$ and $20.2~M_\odot$ mark the corresponding threshold values of~$\lambda$ for the BHs in GW151226. Observe that because these two values are largely similar, the orange curve exhibits just one ``valley'' wherein $p(\log\lambda|d)$ dips below its maximum value. That this dip never goes below 50\% of the maximum is a reflection of the fact that neither spin in GW151226 is well constrained. Nevertheless, we see from the combined posterior in Fig.~\ref{fig:posterior_combined} that events of this kind can still give us some amount of information about the probability of different values of~$\lambda$, although such results must be read with a certain degree of caution, as they are invariably sensitive to our choice of spin priors. What \emph{is} robust about Fig.~\ref{fig:posterior_combined} is the deep valley centered around ${\lambda \sim 73~M_\odot}$, which (as~we discussed) is the result of there being no evidence for a scalarized primary~in~GW190814.

We now come to discuss how the probability density functions in Fig.~\ref{fig:posterior_combined} are normalized. Recall that while we have set the prior on $\lambda$ to be log~uniform with compact support only in the range ${\lambda \in [1, 10^3]~M_\odot}$, this was done purely for the sake of computational efficiency. In reality, this range\,---\,call it $[\lambda_\text{min}, \lambda_\text{max}]$, say\,---\,ought to be much larger. In the extreme case, we might set the lower bound $\lambda_\text{min}$ equal to zero, while the upper bound $\lambda_\text{max}$ can be identified with the very large scale at which Eq.~\eqref{eq:stgb_action} ceases to be a valid effective field theory. Since astrophysical BHs in this theory become indistinguishable from their GR counterparts when~$\lambda$ is too small or too large, the far ends of $p(\log\lambda|d)$ are directly proportional to $\pi(\log\lambda)$. What we therefore have is a posterior that is essentially flat for most points along the $\log\lambda$~line, save for those values of $\lambda$ that are within several orders of magnitude of a solar mass. Statistical measures like the moments of $p(\log\lambda|d)$ are completely dominated by these flat regions, and so carry very little information that would be meaningful to us. All that really matters, in this case, is how the relative value of $p(\log\lambda|d)$ changes as we vary~$\log\lambda$. If $p(\log\lambda|d)$ is found to be close to its maximum value at a given point~$\lambda$, then we learn that the data have no constraining power in this region, whereas if $p(\log\lambda|d)$ is found to be significantly smaller than the maximum, it follows that the data have levied a strong constraint. To emphasize this point visually, we have normalized the probability density functions in Fig.~\ref{fig:posterior_combined} such that they all plateau towards the same maximum value as ${\log_{10}(\lambda/M_\odot) \to 0}$.\footnote{This normalization scheme is actually consistent with requiring that the integral ${\int p(\log\lambda|d)\,\dx\log\lambda = 1}$, since the ``valleys'' in the range ${\lambda \in [1, 10^3]~M_\odot}$ are insignificant compared to the flat regions that extend towards $\lambda_\text{min}$ and $\lambda_\text{max}$. Note, however, that we cannot set~$\lambda_\text{min}$ exactly equal to zero if the above integral is~to~be~finite.
\looseness=-1}

To close this section, it would be nice to quantify what we mean when we say that a value of~$\lambda$ is ``strongly disfavored.'' We do so by computing the Bayes factor
\begin{equation}
	B(\log\lambda)
	=
	\lim_{\lambda_{\min} \to\, 0 } \frac{ p(\log\lambda | d) }{ p(\log\lambda_\text{min}|d)}.
\end{equation}
The denominator $p(\log\lambda_\text{min}|d)$ is equal to the maximum value of our nearly flat posterior, while the particular point ${\lambda \to 0}$ coincides with GR. This tells us that if $p(\log\lambda|d)$ evaluates to a fraction~$B$ of its maximum value, then the corresponding value of~$\lambda$ is $B^{-1}$ times less likely than GR is at being an accurate description of the data. Given this diagnostic, we find that values of ${\lambda \in [56,96]~M_\odot}$ are strongly disfavored~with~${ B \leq 0.1}$.
\looseness=-1

\section{Discussion}
\label{sec:discussion}

The amount of GW transient data available is steadily increasing, with three catalogs having now been released by the LIGO Scientific, Virgo, and KAGRA collaborations \cite{gwtc1, gwtc2, gwtc3}. As~such, it has become both timely and interesting to assess the extent to which these data are useful for probing the existence of new physics beyond~GR.
\looseness=-1

While a dearth of numerical relativity simulations in modified gravity still hinders our ability to model the merger and ringdown phases of a binary's coalescence accurately, in some theories\,---\,namely, scalar-tensor theories\,---\,extensive results from post-Newtonian calculations allow us to construct reliable waveform models that can be used to perform parameter estimation on the inspiral portion of a GW~signal.

In this paper, we confronted this type of waveform model with two real events, GW190814 and GW151226, to establish novel constraints on a particular STGB theory that allows for the spontaneous scalarization of~BHs. Because slowly rotating BHs in this theory must scalarize if their size is comparable to the new length scale~$\lambda$ that is introduced, whereas rapidly rotating BHs of any mass are effectively indistinguishable from~Kerr, we found that GW190814\,---\,whose primary has a spin that is bounded to be small, and whose signal is by all metrics consistent with~GR\,---\,was able to rule out values of~$\lambda$ in a narrow range slightly above~$2M_1$, with $M_1$ denoting the mass of the primary~BH. The other event, GW151226, did not levy any meaningful constraint, but served to illustrate that some amount of information can still be extracted when the spins of both components in a binary are~poorly~measured.

It is worth emphasizing that although we focused on just one particular STGB theory in this study, namely, the one whose coupling function~$f(\phi)$ is given in Eq.~\eqref{eq:stgb_coupling_function}, the universal nature of the \emph{onset} of spontaneous scalarization implies that our general conclusion\,---\,that GW190814 excludes a range values of~$\lambda$ near~$2M_1$\,---\,holds for any STGB theory in this class. The precise range of values that are excluded, however, is model dependent, as it is determined by the particular way in which the BH's scalar charge~$Q$ varies as a function of its mass and spin. For the specific theory in Eq.~\eqref{eq:stgb_coupling_function}, we find that values of~${\lambda \in [56,96]~M_\odot}$ are strongly disfavored with a Bayes~factor~of~$0.1$~or~less.
\looseness=-1

Putting BHs aside, neutron stars are also capable of undergoing spontaneous scalarization in STGB theories~\cite{Silva:2017uqg, Doneva:2017duq, Ventagli:2021ubn}. Indeed, the authors of Ref.~\cite{Danchev:2021tew} used observations of binary pulsars to establish constraints on~$\lambda$ for the same specific theory that we consider, albeit for values of ${\beta \sim 10^4}$. Smaller values of~$\beta$ are challenging to probe due to numerical difficulties encountered when solving the field equations for a neutron star, although an extrapolation of the aforementioned results down to our value of ${\beta = 6}$ suggests that binary pulsars should impose more stringent constraints on~$\lambda$ than~those derived herein. Even~so, our analysis still provides a complementary constraint on~$\lambda$ that was obtained via independent means, i.e., through GW~observations~of~BHs.

In the future, it would certainly be interesting to revisit this analysis when subsequent observing runs uncover more GW~events in which at least one of the two spins is well constrained and found to be small. Alternatively, a possible twist on this study would be to consider the case of spin-induced scalarization~\cite{Dima:2020yac, Herdeiro:2020wei, Berti:2020kgk, Doneva:2020kfv, Hod:2022hfm}. In this variation of the phenomenon, which occurs in STGB theories with a coupling function that satisfies ${f'(0)=0}$ and ${f''(0) < 0}$ [cf.~the discussion above~Eq.~\eqref{eq:stgb_coupling_function}], rapidly rotating BHs are the ones that scalarize if their size is comparable to the length scale~$\lambda$, while slowly rotating BHs remain identical to~Kerr. We expect that a similar analysis could be used to derive meaningful constraints on this class of theories if one or more GW events with a rapidly spinning component are detected.

\acknowledgments
This research was supported by the Center for Research and Development in Mathematics and Applications (CIDMA) through the Portuguese Foundation for Science and Technology (FCT---Funda\c{c}\~ao para a Ci\^encia e a Tecnologia) references UIDB/04106/2020 and UIDP/04106/2020, and the projects PTDC/FIS-OUT/28407/2017, CERN/FIS-PAR/0027/2019, PTDC/FIS-AST/3041/2020, and CERN/FIS-PAR/0024/2021.
We further acknowledge support from the European Union's Horizon 2020 research and innovation (RISE) program H2020-MSCA-RISE-2017 Grant~No.~FunFiCO-777740.
This research has made use of data or software obtained from the Gravitational Wave Open Science~Center~\cite{gwosc}, a~service of LIGO~Laboratory, the LIGO Scientific Collaboration, the Virgo Collaboration, and~KAGRA. LIGO Laboratory and Advanced LIGO are funded by the United States National Science Foundation (NSF) as well as the Science and Technology Facilities Council (STFC) of the United Kingdom, the Max-Planck-Society (MPS), and the State of Niedersachsen/Germany for support of the construction of Advanced LIGO and construction and operation of the GEO600 detector. Additional support for Advanced LIGO was provided by the Australian Research Council. Virgo is funded, through the European Gravitational Observatory (EGO), by the French Centre National de Recherche Scientifique (CNRS), the Italian Istituto Nazionale di Fisica Nucleare (INFN) and the Dutch Nikhef, with contributions by institutions from Belgium, Germany, Greece, Hungary, Ireland, Japan, Monaco, Poland, Portugal, and Spain. The construction and operation of KAGRA are funded by the Ministry of Education, Culture, Sports, Science and Technology (MEXT) and the Japan Society for the Promotion of Science (JSPS), the National Research Foundation (NRF) and Ministry of Science and ICT (MSIT) in Korea, Academia Sinica (AS) and the Ministry of Science and Technology (MoST) in Taiwan.
This research has also made use of the Astropy~\cite{astropy:2013, astropy:2018}, Bilby~\cite{Ashton:2018jfp, Romero-Shaw:2020owr}, Corner~\cite{corner}, Dynesty~\cite{dynesty}, GWpy~\cite{gwpy}, LALSuite~\cite{lalsuite}, Matplotlib~\cite{matplotlib}, and SciPy~\cite{scipy} software packages.
\looseness=-1

\appendix*
\section{Correlations\protect\\between the source parameters}

To complement Fig.~\ref{fig:posterior_GW190814} in the main text, Fig.~\ref{fig:posterior_GW190814_extended} presents a larger corner plot of the posterior distribution for~GW190814. The five parameters we focus on are the coupling constant~$\lambda$, the component masses $M_1$ and $M_2$ in the rest frame of the source, and the dimensionless spin magnitudes $\chi_1$ and $\chi_2$. The blue shaded regions represent the posterior under the assumption of the STGB model, whereas the orange lines represent the corresponding posterior under the assumption of~GR. Parameter estimation samples for the latter were obtained directly~from~Ref.~\cite{dccGW190814}.

Notice in Fig.~\ref{fig:posterior_GW190814_extended} that the source parameters $\{ M_1, M_2, \chi_1, \chi_2 \}$ have much larger uncertainties in the STGB model than in~GR. Moreover, notice that the one-dimensional posteriors for the first three of these parameters have become bimodal in the STGB case. What is interesting about these observations is that they are not due to extra degeneracies introduced by the inclusion of the additional parameter~$\lambda$ (a~perfectly reasonable first guess), but are actually the result of us having limited our analysis to the inspiral stage. We verified that this was the main cause by sampling the posterior for the GR case ourselves while using only the inspiral portion of the signal. (The~parameter estimation samples from Ref.~\cite{dccGW190814} were obtained using the full signal, including merger~and~ringdown.) That these differences between the two cases are driven by this truncation at the end of the inspiral\,---\,and not by the inclusion of~$\lambda$\,---\,is good news, as it suggests that our result for the marginalized posterior 
${p(\vartheta|d) = \int p(\log\lambda, \vartheta|d) \,\dx\log\lambda}$
is mostly robust to changes in the domain of~$\lambda$, for instance, if we were to extend the log-uniform prior on $\lambda$ to have support over the full range~$[\lambda_\text{min}, \lambda_\text{max}]$ (see~the discussion in Sec.~\ref{sec:gw_results}). The only exception to this is the marginalized distribution of~$\chi_2$, which will lose its preference for higher spins as we sample more points with smaller or larger values of~$\lambda$, i.e., points in regions of the parameter space that are indistinguishable~from~GR.

We conclude with a remark on the distributions of $M_1$, $M_2$, and $\chi_1$. While the bimodal behavior in Fig.~\ref{fig:posterior_GW190814_extended} has little impact on the maximum \emph{a posteriori} values of the masses, larger uncertainties mean a substantially weaker upper bound on~$\chi_1$: we find a 90\% upper limit of ${\chi_1 < 0.35}$ in the STGB case, compared to ${\chi_1 < 0.07}$ in~GR. This weaker upper bound is nevertheless still small enough to ensure that, for this particular theory, the primary of GW190814 inevitably scalarizes for some range of values of~$\lambda$.

\bibliography{main}
\end{document}

%% file: stylesheet.tex
\usepackage[stretch=10,final]{microtype}
\usepackage{nowidow}
\linespread{1}
\setlength{\parskip}{0pt}

\let\revtitle\maketitle
\renewcommand{\maketitle}{%
	\revtitle
	\tolerance=200
	\hyphenpenalty=200
}

\setlength{\textfloatsep}{0.7em}
\setlength{\abovecaptionskip}{5pt}





\DeclareSymbolFont{cmreg}{OT1}{cmr}{m}{n}
\DeclareSymbolFont{cmmath}{OML}{cmm}{m}{i}
\DeclareSymbolFont{cmsymbols}{OMS}{cmsy}{m}{n}
\DeclareSymbolFont{cmlargesymbols}{OMX}{cmex}{m}{n}
\DeclareSymbolFontAlphabet{\mathcal}{cmsymbols}

\DeclareMathSymbol{\partial}{0}{cmmath}{64}
\DeclareMathSymbol{g}{\mathalpha}{cmmath}{103}
\DeclareMathSymbol{\eta}{0}{cmmath}{17}
\DeclareMathSymbol{\kappa}{0}{cmmath}{20}
\DeclareMathSymbol{\mu}{0}{cmmath}{22}
\DeclareMathSymbol{\nu}{0}{cmmath}{23}
\DeclareMathSymbol{\rho}{0}{cmmath}{26}
\DeclareMathSymbol{\sigma}{0}{cmmath}{27}
\DeclareMathSymbol{\ell}{0}{cmmath}{96}

\DeclareMathSymbol{\ointop}{\mathop}{cmlargesymbols}{72}
\DeclareMathSymbol{\intop}{\mathop}{cmlargesymbols}{82}
\DeclareMathDelimiter{(}{\mathopen}{cmreg}{40}{cmlargesymbols}{0}
\DeclareMathDelimiter{)}{\mathclose}{cmreg}{41}{cmlargesymbols}{1}

\DeclareMathDelimiter{(}{\mathopen}{cmreg}{40}{cmlargesymbols}{0}
\DeclareMathDelimiter{)}{\mathclose}{cmreg}{41}{cmlargesymbols}{1}
\DeclareMathDelimiter{[}{\mathopen}{cmreg}{91}{cmlargesymbols}{2}
\DeclareMathDelimiter{]}{\mathclose}{cmreg}{93}{cmlargesymbols}{3}


\usepackage{titlesec}

\newcommand{\mysectionnumbering}{\thesection.~}

\titleformat{\section}{\bfseries\center\uppercase}{\mysectionnumbering}{0em}{}
\titleformat{\subsection}{\bfseries\center}{}{0.1em}{\thesubsection.~}
\titleformat{\subsubsection}{\bfseries\itshape\center}{}{0.1em}{\thesubsubsection.~}

\titlespacing{\section}{0pt}{1.7em plus 0.9em minus 0.2em}{0.7em plus 0.2em minus 0em}
\titlespacing{\subsection}{0pt}{1.5em plus 0.1em minus 0.1em}{0.5em}
\titlespacing{\subsubsection}{0pt}{1.5em plus 0.1em minus 0.1em}{0.5em}

\titleformat{\paragraph}[runin]{\itshape}{}{0em}{}[.---]
\titlespacing{\paragraph}{\the\parindent}{0em}{0em}

\let\oldappendix\appendix

\usepackage{etoolbox}
\makeatletter
\renewcommand{\appendix}{\@ifstar\oneappendix\manyappendices}
\makeatother

\newcommand{\defappsec}{%
	\let\oldsection\section
	\renewcommand{\section}[1]{%
	\ifstrempty{##1}
		{\oldsection{}}
		{\oldsection{:~{##1}}}
	}
}

\newcommand{\oneappendix}{%
\oldappendix*
\defappsec
\renewcommand{\mysectionnumbering}{\MakeUppercase{Appendix}}
\renewcommand\theequation{\Alph{section}\arabic{equation}}
}
\newcommand{\manyappendices}{%
\oldappendix
\defappsec
\renewcommand{\mysectionnumbering}{\MakeUppercase{Appendix}~\thesection}
\renewcommand\theequation{\Alph{section}\arabic{equation}}
}


\makeatletter
\renewcommand\@makefntext[1]{%
\noindent{\hspace{1em}}{\@makefnmark}#1}
\makeatother

\makeatletter
\renewcommand\@makefnmark{\hbox{\color{black}\@textsuperscript{\normalfont\@thefnmark}}}
\makeatother



\setlength{\skip\footins}{15pt}

\renewcommand{\footnoterule}{%
  \kern -3pt
  \hrule width 1.2cm
  \kern 4pt
}


\usepackage[dvipsnames]{xcolor}
\definecolor{revblue}{HTML}{2d3092}
\colorlet{blue}{revblue}

\let\revcite\cite
\renewcommand\cite[1]{\mbox{\color{blue}\revcite{#1}}}

\let\reveqref\eqref
\renewcommand\eqref[1]{{\color{blue}\reveqref{#1}}}

\IfFileExists{apsrev-custom.bst}{\bibliographystyle{apsrev-custom}}{}


\newlength\mycitespacing
\settowidth\mycitespacing{\space}
\addtolength\mycitespacing{-0.5pt}
\setcitestyle{citesep={,\kern-\mycitespacing}}


%

%% file: main.bbl
\begin{thebibliography}{108}%
\makeatletter
\providecommand \@ifxundefined [1]{%
 \@ifx{#1\undefined}
}%
\providecommand \@ifnum [1]{%
 \ifnum #1\expandafter \@firstoftwo
 \else \expandafter \@secondoftwo
 \fi
}%
\providecommand \@ifx [1]{%
 \ifx #1\expandafter \@firstoftwo
 \else \expandafter \@secondoftwo
 \fi
}%
\providecommand \natexlab [1]{#1}%
\providecommand \enquote  [1]{``#1''}%
\providecommand \bibnamefont  [1]{#1}%
\providecommand \bibfnamefont [1]{#1}%
\providecommand \citenamefont [1]{#1}%
\providecommand \href@noop [0]{\@secondoftwo}%
\providecommand \href [0]{\begingroup \@sanitize@url \@href}%
\providecommand \@href[1]{\@@startlink{#1}\@@href}%
\providecommand \@@href[1]{\endgroup#1\@@endlink}%
\providecommand \@sanitize@url [0]{\catcode `\\12\catcode `\$12\catcode
  `\&12\catcode `\#12\catcode `\^12\catcode `\_12\catcode `\%12\relax}%
\providecommand \@@startlink[1]{}%
\providecommand \@@endlink[0]{}%
\providecommand \url  [0]{\begingroup\@sanitize@url \@url }%
\providecommand \@url [1]{\endgroup\@href {#1}{\urlprefix }}%
\providecommand \urlprefix  [0]{URL }%
\providecommand \Eprint [0]{\href }%
\providecommand \doibase [0]{https://doi.org/}%
\providecommand \selectlanguage [0]{\@gobble}%
\providecommand \bibinfo  [0]{\@secondoftwo}%
\providecommand \bibfield  [0]{\@secondoftwo}%
\providecommand \translation [1]{[#1]}%
\providecommand \BibitemOpen [0]{}%
\providecommand \bibitemStop [0]{}%
\providecommand \bibitemNoStop [0]{.\EOS\space}%
\providecommand \EOS [0]{\spacefactor3000\relax}%
\providecommand \BibitemShut  [1]{\csname bibitem#1\endcsname}%
\let\auto@bib@innerbib\@empty
\bibitem [{\citenamefont {Will}(2014)}]{Will:2014kxa}%
  \BibitemOpen
  \bibfield  {author} {\bibinfo {author} {\bibfnamefont {C.~M.}\ \bibnamefont
  {Will}},\ }\bibfield  {title} {\bibinfo {title} {The confrontation between
  general relativity and experiment},\ }\href
  {https://doi.org/10.12942/lrr-2014-4} {\bibfield  {journal} {\bibinfo
  {journal} {Living Rev. Relativity}\ }\textbf {\bibinfo {volume} {17}},\
  \bibinfo {pages} {4} (\bibinfo {year} {2014})},\ \Eprint
  {https://arxiv.org/abs/1403.7377} {arXiv:1403.7377}\BibitemShut {NoStop}%
\bibitem [{\citenamefont {Berti}\ \emph {et~al.}(2015)\citenamefont {Berti},
  \citenamefont {Barausse}, \citenamefont {Cardoso}, \citenamefont {Gualtieri},
  \citenamefont {Pani}, \citenamefont {Sperhake}, \citenamefont {Stein},
  \citenamefont {Wex}, \citenamefont {Yagi}, \citenamefont {Baker} \emph
  {et~al.}}]{Berti:2015itd}%
  \BibitemOpen
  \bibfield  {author} {\bibinfo {author} {\bibfnamefont {E.}~\bibnamefont
  {Berti}}, \bibinfo {author} {\bibfnamefont {E.}~\bibnamefont {Barausse}},
  \bibinfo {author} {\bibfnamefont {V.}~\bibnamefont {Cardoso}}, \bibinfo
  {author} {\bibfnamefont {L.}~\bibnamefont {Gualtieri}}, \bibinfo {author}
  {\bibfnamefont {P.}~\bibnamefont {Pani}}, \bibinfo {author} {\bibfnamefont
  {U.}~\bibnamefont {Sperhake}}, \bibinfo {author} {\bibfnamefont {L.~C.}\
  \bibnamefont {Stein}}, \bibinfo {author} {\bibfnamefont {N.}~\bibnamefont
  {Wex}}, \bibinfo {author} {\bibfnamefont {K.}~\bibnamefont {Yagi}}, \bibinfo
  {author} {\bibfnamefont {T.}~\bibnamefont {Baker}},  \emph {et~al.},\
  }\bibfield  {title} {\bibinfo {title} {Testing general relativity with
  present and future astrophysical observations},\ }\href
  {https://doi.org/10.1088/0264-9381/32/24/243001} {\bibfield  {journal}
  {\bibinfo  {journal} {Classical Quantum Gravity}\ }\textbf {\bibinfo {volume}
  {32}},\ \bibinfo {pages} {243001} (\bibinfo {year} {2015})},\ \Eprint
  {https://arxiv.org/abs/1501.07274} {arXiv:1501.07274}\BibitemShut {NoStop}%
\bibitem [{\citenamefont {Yunes}\ \emph {et~al.}(2016)\citenamefont {Yunes},
  \citenamefont {Yagi},\  and\ \citenamefont {Pretorius}}]{Yunes:2016jcc}%
  \BibitemOpen
  \bibfield  {author} {\bibinfo {author} {\bibfnamefont {N.}~\bibnamefont
  {Yunes}}, \bibinfo {author} {\bibfnamefont {K.}~\bibnamefont {Yagi}},  and\
  \bibinfo {author} {\bibfnamefont {F.}~\bibnamefont {Pretorius}},\ }\bibfield
  {title} {\bibinfo {title} {{Theoretical physics implications of the binary
  black-hole mergers GW150914 and GW151226}},\ }\href
  {https://doi.org/10.1103/PhysRevD.94.084002} {\bibfield  {journal} {\bibinfo
  {journal} {Phys. Rev. D}\ }\textbf {\bibinfo {volume} {94}},\ \bibinfo
  {pages} {084002} (\bibinfo {year} {2016})},\ \Eprint
  {https://arxiv.org/abs/1603.08955} {arXiv:1603.08955}\BibitemShut {NoStop}%
\bibitem [{\citenamefont {Damour}()}]{Damour:PDGReview}%
  \BibitemOpen
  \bibfield  {author} {\bibinfo {author} {\bibfnamefont {T.}~\bibnamefont
  {Damour}},\ }\href@noop {} {\bibinfo {title} {Experimental tests of
  gravitational theory}},\ \bibinfo {note} {{in Review of Particle Physics, P.
  A. Zyla \emph{et al.} (Particle Data Group),
  \href{https://doi.org/10.1093/ptep/ptaa104}{Prog. Theor. Exp. Phys.
  \textbf{2020}, 083C01 (2020) and 2021 update}.}}\BibitemShut {Stop}%
\bibitem [{\citenamefont {Ostrogradsky}(1850)}]{Ostrogradsky:1850fid}%
  \BibitemOpen
  \bibfield  {author} {\bibinfo {author} {\bibfnamefont {M.}~\bibnamefont
  {Ostrogradsky}},\ }\bibfield  {title} {\bibinfo {title} {{M\'emoires sur les
  \'equations diff\'erentielles, relatives au probl\`eme des
  isop\'erim\`etres}},\ }\href@noop {} {\bibfield  {journal} {\bibinfo
  {journal} {Mem. Acad. St. Petersbourg}\ }\textbf {\bibinfo {volume} {6}},\
  \bibinfo {pages} {385} (\bibinfo {year} {1850})}\BibitemShut {NoStop}%
\bibitem [{\citenamefont {Lovelock}(1971)}]{Lovelock:1971yv}%
  \BibitemOpen
  \bibfield  {author} {\bibinfo {author} {\bibfnamefont {D.}~\bibnamefont
  {Lovelock}},\ }\bibfield  {title} {\bibinfo {title} {{The Einstein tensor and
  its generalizations}},\ }\href {https://doi.org/10.1063/1.1665613} {\bibfield
   {journal} {\bibinfo  {journal} {J. Math. Phys. (N.Y.)}\ }\textbf {\bibinfo
  {volume} {12}},\ \bibinfo {pages} {498} (\bibinfo {year} {1971})}\BibitemShut
  {NoStop}%
\bibitem [{\citenamefont {Zwiebach}(1985)}]{Zwiebach:1985uq}%
  \BibitemOpen
  \bibfield  {author} {\bibinfo {author} {\bibfnamefont {B.}~\bibnamefont
  {Zwiebach}},\ }\bibfield  {title} {\bibinfo {title} {Curvature squared terms
  and string theories},\ }\href {https://doi.org/10.1016/0370-2693(85)91616-8}
  {\bibfield  {journal} {\bibinfo  {journal} {Phys. Lett.}\ }\textbf {\bibinfo
  {volume} {156B}},\ \bibinfo {pages} {315} (\bibinfo {year}
  {1985})}\BibitemShut {NoStop}%
\bibitem [{\citenamefont {Kanti}\ \emph {et~al.}(1996)\citenamefont {Kanti},
  \citenamefont {Mavromatos}, \citenamefont {Rizos}, \citenamefont {Tamvakis},\
   and\ \citenamefont {Winstanley}}]{Kanti:1995vq}%
  \BibitemOpen
  \bibfield  {author} {\bibinfo {author} {\bibfnamefont {P.}~\bibnamefont
  {Kanti}}, \bibinfo {author} {\bibfnamefont {N.~E.}\ \bibnamefont
  {Mavromatos}}, \bibinfo {author} {\bibfnamefont {J.}~\bibnamefont {Rizos}},
  \bibinfo {author} {\bibfnamefont {K.}~\bibnamefont {Tamvakis}},  and\
  \bibinfo {author} {\bibfnamefont {E.}~\bibnamefont {Winstanley}},\ }\bibfield
   {title} {\bibinfo {title} {{Dilatonic black holes in higher curvature string
  gravity}},\ }\href {https://doi.org/10.1103/PhysRevD.54.5049} {\bibfield
  {journal} {\bibinfo  {journal} {Phys. Rev. D}\ }\textbf {\bibinfo {volume}
  {54}},\ \bibinfo {pages} {5049} (\bibinfo {year} {1996})},\ \Eprint
  {https://arxiv.org/abs/hep-th/9511071} {arXiv:hep-th/9511071}\BibitemShut
  {NoStop}%
\bibitem [{\citenamefont {Kleihaus}\ \emph {et~al.}(2011)\citenamefont
  {Kleihaus}, \citenamefont {Kunz},\  and\ \citenamefont
  {Radu}}]{Kleihaus:2011tg}%
  \BibitemOpen
  \bibfield  {author} {\bibinfo {author} {\bibfnamefont {B.}~\bibnamefont
  {Kleihaus}}, \bibinfo {author} {\bibfnamefont {J.}~\bibnamefont {Kunz}},
  and\ \bibinfo {author} {\bibfnamefont {E.}~\bibnamefont {Radu}},\ }\bibfield
  {title} {\bibinfo {title} {{Rotating Black Holes in Dilatonic
  Einstein-Gauss-Bonnet Theory}},\ }\href
  {https://doi.org/10.1103/PhysRevLett.106.151104} {\bibfield  {journal}
  {\bibinfo  {journal} {Phys. Rev. Lett.}\ }\textbf {\bibinfo {volume} {106}},\
  \bibinfo {pages} {151104} (\bibinfo {year} {2011})},\ \Eprint
  {https://arxiv.org/abs/1101.2868} {arXiv:1101.2868}\BibitemShut {NoStop}%
\bibitem [{\citenamefont {Sotiriou}\  and\ \citenamefont
  {Zhou}(2014)}]{Sotiriou:2014pfa}%
  \BibitemOpen
  \bibfield  {author} {\bibinfo {author} {\bibfnamefont {T.~P.}\ \bibnamefont
  {Sotiriou}} and\ \bibinfo {author} {\bibfnamefont {S.-Y.}\ \bibnamefont
  {Zhou}},\ }\bibfield  {title} {\bibinfo {title} {{Black hole hair in
  generalized scalar-tensor gravity: An explicit example}},\ }\href
  {https://doi.org/10.1103/PhysRevD.90.124063} {\bibfield  {journal} {\bibinfo
  {journal} {Phys. Rev. D}\ }\textbf {\bibinfo {volume} {90}},\ \bibinfo
  {pages} {124063} (\bibinfo {year} {2014})},\ \Eprint
  {https://arxiv.org/abs/1408.1698} {arXiv:1408.1698}\BibitemShut {NoStop}%
\bibitem [{\citenamefont {Delgado}\ \emph {et~al.}(2020)\citenamefont
  {Delgado}, \citenamefont {Herdeiro},\  and\ \citenamefont
  {Radu}}]{Delgado:2020rev}%
  \BibitemOpen
  \bibfield  {author} {\bibinfo {author} {\bibfnamefont {J.~F.~M.}\
  \bibnamefont {Delgado}}, \bibinfo {author} {\bibfnamefont {C.~A.~R.}\
  \bibnamefont {Herdeiro}},  and\ \bibinfo {author} {\bibfnamefont
  {E.}~\bibnamefont {Radu}},\ }\bibfield  {title} {\bibinfo {title} {{Spinning
  black holes in shift-symmetric Horndeski theory}},\ }\href
  {https://doi.org/10.1007/JHEP04(2020)180} {\bibfield  {journal} {\bibinfo
  {journal} {J. High Energy Phys.}\ }04 (2020)\ 180},\ \Eprint
  {https://arxiv.org/abs/2002.05012} {arXiv:2002.05012}\BibitemShut {NoStop}%
\bibitem [{\citenamefont {Herdeiro}\  and\ \citenamefont
  {Radu}(2015)}]{Herdeiro:2015waa}%
  \BibitemOpen
  \bibfield  {author} {\bibinfo {author} {\bibfnamefont {C.~A.~R.}\
  \bibnamefont {Herdeiro}} and\ \bibinfo {author} {\bibfnamefont
  {E.}~\bibnamefont {Radu}},\ }\bibfield  {title} {\bibinfo {title}
  {{Asymptotically flat black holes with scalar hair: A review}},\ }\href
  {https://doi.org/10.1142/S0218271815420146} {\bibfield  {journal} {\bibinfo
  {journal} {Int. J. Mod. Phys. D}\ }\textbf {\bibinfo {volume} {24}},\
  \bibinfo {pages} {1542014} (\bibinfo {year} {2015})},\ \Eprint
  {https://arxiv.org/abs/1504.08209} {arXiv:1504.08209}\BibitemShut {NoStop}%
\bibitem [{\citenamefont {Doneva}\  and\ \citenamefont
  {Yazadjiev}(2018{\natexlab{a}})}]{Doneva:2017bvd}%
  \BibitemOpen
  \bibfield  {author} {\bibinfo {author} {\bibfnamefont {D.~D.}\ \bibnamefont
  {Doneva}} and\ \bibinfo {author} {\bibfnamefont {S.~S.}\ \bibnamefont
  {Yazadjiev}},\ }\bibfield  {title} {\bibinfo {title} {{New Gauss-Bonnet Black
  Holes with Curvature-Induced Scalarization in Extended Scalar-Tensor
  Theories}},\ }\href {https://doi.org/10.1103/PhysRevLett.120.131103}
  {\bibfield  {journal} {\bibinfo  {journal} {Phys. Rev. Lett.}\ }\textbf
  {\bibinfo {volume} {120}},\ \bibinfo {pages} {131103} (\bibinfo {year}
  {2018}{\natexlab{a}})},\ \Eprint {https://arxiv.org/abs/1711.01187}
  {arXiv:1711.01187}\BibitemShut {NoStop}%
\bibitem [{\citenamefont {Silva}\ \emph {et~al.}(2018)\citenamefont {Silva},
  \citenamefont {Sakstein}, \citenamefont {Gualtieri}, \citenamefont
  {Sotiriou},\  and\ \citenamefont {Berti}}]{Silva:2017uqg}%
  \BibitemOpen
  \bibfield  {author} {\bibinfo {author} {\bibfnamefont {H.~O.}\ \bibnamefont
  {Silva}}, \bibinfo {author} {\bibfnamefont {J.}~\bibnamefont {Sakstein}},
  \bibinfo {author} {\bibfnamefont {L.}~\bibnamefont {Gualtieri}}, \bibinfo
  {author} {\bibfnamefont {T.~P.}\ \bibnamefont {Sotiriou}},  and\ \bibinfo
  {author} {\bibfnamefont {E.}~\bibnamefont {Berti}},\ }\bibfield  {title}
  {\bibinfo {title} {{Spontaneous Scalarization of Black Holes and Compact
  Stars from a Gauss-Bonnet Coupling}},\ }\href
  {https://doi.org/10.1103/PhysRevLett.120.131104} {\bibfield  {journal}
  {\bibinfo  {journal} {Phys. Rev. Lett.}\ }\textbf {\bibinfo {volume} {120}},\
  \bibinfo {pages} {131104} (\bibinfo {year} {2018})},\ \Eprint
  {https://arxiv.org/abs/1711.02080} {arXiv:1711.02080}\BibitemShut {NoStop}%
\bibitem [{\citenamefont {Antoniou}\ \emph {et~al.}(2018)\citenamefont
  {Antoniou}, \citenamefont {Bakopoulos},\  and\ \citenamefont
  {Kanti}}]{Antoniou:2017acq}%
  \BibitemOpen
  \bibfield  {author} {\bibinfo {author} {\bibfnamefont {G.}~\bibnamefont
  {Antoniou}}, \bibinfo {author} {\bibfnamefont {A.}~\bibnamefont
  {Bakopoulos}},  and\ \bibinfo {author} {\bibfnamefont {P.}~\bibnamefont
  {Kanti}},\ }\bibfield  {title} {\bibinfo {title} {{Evasion of No-Hair
  Theorems and Novel Black-Hole Solutions in Gauss-Bonnet Theories}},\ }\href
  {https://doi.org/10.1103/PhysRevLett.120.131102} {\bibfield  {journal}
  {\bibinfo  {journal} {Phys. Rev. Lett.}\ }\textbf {\bibinfo {volume} {120}},\
  \bibinfo {pages} {131102} (\bibinfo {year} {2018})},\ \Eprint
  {https://arxiv.org/abs/1711.03390} {arXiv:1711.03390}\BibitemShut {NoStop}%
\bibitem [{\citenamefont {Damour}\  and\ \citenamefont
  {Esposito-Far\`ese}(1993)}]{Damour:1993hw}%
  \BibitemOpen
  \bibfield  {author} {\bibinfo {author} {\bibfnamefont {T.}~\bibnamefont
  {Damour}} and\ \bibinfo {author} {\bibfnamefont {G.}~\bibnamefont
  {Esposito-Far\`ese}},\ }\bibfield  {title} {\bibinfo {title}
  {{Nonperturbative Strong-Field Effects in Tensor-Scalar Theories of
  Gravitation}},\ }\href {https://doi.org/10.1103/PhysRevLett.70.2220}
  {\bibfield  {journal} {\bibinfo  {journal} {Phys. Rev. Lett.}\ }\textbf
  {\bibinfo {volume} {70}},\ \bibinfo {pages} {2220} (\bibinfo {year}
  {1993})}\BibitemShut {NoStop}%
\bibitem [{\citenamefont {Wang}\ \emph {et~al.}(2021)\citenamefont {Wang},
  \citenamefont {Tang}, \citenamefont {Li}, \citenamefont {Han},\  and\
  \citenamefont {Fan}}]{Wang:2021yll}%
  \BibitemOpen
  \bibfield  {author} {\bibinfo {author} {\bibfnamefont {H.-T.}\ \bibnamefont
  {Wang}}, \bibinfo {author} {\bibfnamefont {S.-P.}\ \bibnamefont {Tang}},
  \bibinfo {author} {\bibfnamefont {P.-C.}\ \bibnamefont {Li}}, \bibinfo
  {author} {\bibfnamefont {M.-Z.}\ \bibnamefont {Han}},  and\ \bibinfo {author}
  {\bibfnamefont {Y.-Z.}\ \bibnamefont {Fan}},\ }\bibfield  {title} {\bibinfo
  {title} {{Tight constraints on Einstein-dilation-Gauss-Bonnet gravity from
  GW190412 and GW190814}},\ }\href
  {https://doi.org/10.1103/PhysRevD.104.024015} {\bibfield  {journal} {\bibinfo
   {journal} {Phys. Rev. D}\ }\textbf {\bibinfo {volume} {104}},\ \bibinfo
  {pages} {024015} (\bibinfo {year} {2021})},\ \Eprint
  {https://arxiv.org/abs/2104.07590} {arXiv:2104.07590}\BibitemShut {NoStop}%
\bibitem [{\citenamefont {Perkins}\ \emph {et~al.}(2021)\citenamefont
  {Perkins}, \citenamefont {Nair}, \citenamefont {Silva},\  and\ \citenamefont
  {Yunes}}]{Perkins:2021mhb}%
  \BibitemOpen
  \bibfield  {author} {\bibinfo {author} {\bibfnamefont {S.~E.}\ \bibnamefont
  {Perkins}}, \bibinfo {author} {\bibfnamefont {R.}~\bibnamefont {Nair}},
  \bibinfo {author} {\bibfnamefont {H.~O.}\ \bibnamefont {Silva}},  and\
  \bibinfo {author} {\bibfnamefont {N.}~\bibnamefont {Yunes}},\ }\bibfield
  {title} {\bibinfo {title} {{Improved gravitational-wave constraints on
  higher-order curvature theories of gravity}},\ }\href
  {https://doi.org/10.1103/PhysRevD.104.024060} {\bibfield  {journal} {\bibinfo
   {journal} {Phys. Rev. D}\ }\textbf {\bibinfo {volume} {104}},\ \bibinfo
  {pages} {024060} (\bibinfo {year} {2021})},\ \Eprint
  {https://arxiv.org/abs/2104.11189} {arXiv:2104.11189}\BibitemShut {NoStop}%
\bibitem [{\citenamefont {Lyu}\ \emph {et~al.}(2022)\citenamefont {Lyu},
  \citenamefont {Jiang},\  and\ \citenamefont {Yagi}}]{Lyu:2022gdr}%
  \BibitemOpen
  \bibfield  {author} {\bibinfo {author} {\bibfnamefont {Z.}~\bibnamefont
  {Lyu}}, \bibinfo {author} {\bibfnamefont {N.}~\bibnamefont {Jiang}},  and\
  \bibinfo {author} {\bibfnamefont {K.}~\bibnamefont {Yagi}},\ }\bibfield
  {title} {\bibinfo {title} {{Constraints on Einstein-dilation-Gauss-Bonnet
  gravity from black hole-neutron star gravitational wave events}},\ }\href
  {https://doi.org/10.1103/PhysRevD.105.064001} {\bibfield  {journal} {\bibinfo
   {journal} {Phys. Rev. D}\ }\textbf {\bibinfo {volume} {105}},\ \bibinfo
  {pages} {064001} (\bibinfo {year} {2022})},\ \Eprint
  {https://arxiv.org/abs/2201.02543} {arXiv:2201.02543}\BibitemShut {NoStop}%
\bibitem [{\citenamefont {Abbott}\ \emph
  {et~al.}(2020{\natexlab{a}})\citenamefont {Abbott} \emph
  {et~al.}}]{LIGOScientific:2020zkf}%
  \BibitemOpen
  \bibfield  {author} {\bibinfo {author} {\bibfnamefont {R.}~\bibnamefont
  {Abbott}} \emph {et~al.} (\bibinfo {collaboration} {LIGO Scientific and Virgo
  Collaborations}),\ }\bibfield  {title} {\bibinfo {title} {{GW190814:
  Gravitational waves from the coalescence of a 23 solar mass black hole with a
  2.6 solar mass compact object}},\ }\href
  {https://doi.org/10.3847/2041-8213/ab960f} {\bibfield  {journal} {\bibinfo
  {journal} {Astrophys. J. Lett.}\ }\textbf {\bibinfo {volume} {896}},\
  \bibinfo {pages} {L44} (\bibinfo {year} {2020}{\natexlab{a}})},\ \Eprint
  {https://arxiv.org/abs/2006.12611} {arXiv:2006.12611}\BibitemShut {NoStop}%
\bibitem [{\citenamefont {Abbott}\ \emph
  {et~al.}(2016{\natexlab{a}})\citenamefont {Abbott} \emph
  {et~al.}}]{LIGOScientific:2016sjg}%
  \BibitemOpen
  \bibfield  {author} {\bibinfo {author} {\bibfnamefont {B.~P.}\ \bibnamefont
  {Abbott}} \emph {et~al.} (\bibinfo {collaboration} {LIGO Scientific and Virgo
  Collaborations}),\ }\bibfield  {title} {\bibinfo {title} {{GW151226:
  Observation of Gravitational Waves from a 22-Solar-Mass Binary Black Hole
  Coalescence}},\ }\href {https://doi.org/10.1103/PhysRevLett.116.241103}
  {\bibfield  {journal} {\bibinfo  {journal} {Phys. Rev. Lett.}\ }\textbf
  {\bibinfo {volume} {116}},\ \bibinfo {pages} {241103} (\bibinfo {year}
  {2016}{\natexlab{a}})},\ \Eprint {https://arxiv.org/abs/1606.04855}
  {arXiv:1606.04855}\BibitemShut {NoStop}%
\bibitem [{\citenamefont {Cunha}\ \emph {et~al.}(2019)\citenamefont {Cunha},
  \citenamefont {Herdeiro},\  and\ \citenamefont {Radu}}]{Cunha:2019dwb}%
  \BibitemOpen
  \bibfield  {author} {\bibinfo {author} {\bibfnamefont {P.~V.~P.}\
  \bibnamefont {Cunha}}, \bibinfo {author} {\bibfnamefont {C.~A.~R.}\
  \bibnamefont {Herdeiro}},  and\ \bibinfo {author} {\bibfnamefont
  {E.}~\bibnamefont {Radu}},\ }\bibfield  {title} {\bibinfo {title}
  {{Spontaneously Scalarized Kerr Black Holes in Extended
  Scalar-Tensor--Gauss-Bonnet Gravity}},\ }\href
  {https://doi.org/10.1103/PhysRevLett.123.011101} {\bibfield  {journal}
  {\bibinfo  {journal} {Phys. Rev. Lett.}\ }\textbf {\bibinfo {volume} {123}},\
  \bibinfo {pages} {011101} (\bibinfo {year} {2019})},\ \Eprint
  {https://arxiv.org/abs/1904.09997} {arXiv:1904.09997}\BibitemShut {NoStop}%
\bibitem [{\citenamefont {Bl\'azquez-Salcedo}\ \emph
  {et~al.}(2018)\citenamefont {Bl\'azquez-Salcedo}, \citenamefont {Doneva},
  \citenamefont {Kunz},\  and\ \citenamefont
  {Yazadjiev}}]{Blazquez-Salcedo:2018jnn}%
  \BibitemOpen
  \bibfield  {author} {\bibinfo {author} {\bibfnamefont {J.~L.}\ \bibnamefont
  {Bl\'azquez-Salcedo}}, \bibinfo {author} {\bibfnamefont {D.~D.}\ \bibnamefont
  {Doneva}}, \bibinfo {author} {\bibfnamefont {J.}~\bibnamefont {Kunz}},  and\
  \bibinfo {author} {\bibfnamefont {S.~S.}\ \bibnamefont {Yazadjiev}},\
  }\bibfield  {title} {\bibinfo {title} {{Radial perturbations of the
  scalarized Einstein-Gauss-Bonnet black holes}},\ }\href
  {https://doi.org/10.1103/PhysRevD.98.084011} {\bibfield  {journal} {\bibinfo
  {journal} {Phys. Rev. D}\ }\textbf {\bibinfo {volume} {98}},\ \bibinfo
  {pages} {084011} (\bibinfo {year} {2018})},\ \Eprint
  {https://arxiv.org/abs/1805.05755} {arXiv:1805.05755}\BibitemShut {NoStop}%
\bibitem [{\citenamefont {Silva}\ \emph {et~al.}(2019)\citenamefont {Silva},
  \citenamefont {Macedo}, \citenamefont {Sotiriou}, \citenamefont {Gualtieri},
  \citenamefont {Sakstein},\  and\ \citenamefont {Berti}}]{Silva:2018qhn}%
  \BibitemOpen
  \bibfield  {author} {\bibinfo {author} {\bibfnamefont {H.~O.}\ \bibnamefont
  {Silva}}, \bibinfo {author} {\bibfnamefont {C.~F.~B.}\ \bibnamefont
  {Macedo}}, \bibinfo {author} {\bibfnamefont {T.~P.}\ \bibnamefont
  {Sotiriou}}, \bibinfo {author} {\bibfnamefont {L.}~\bibnamefont {Gualtieri}},
  \bibinfo {author} {\bibfnamefont {J.}~\bibnamefont {Sakstein}},  and\
  \bibinfo {author} {\bibfnamefont {E.}~\bibnamefont {Berti}},\ }\bibfield
  {title} {\bibinfo {title} {{Stability of scalarized black hole solutions in
  scalar-Gauss-Bonnet gravity}},\ }\href
  {https://doi.org/10.1103/PhysRevD.99.064011} {\bibfield  {journal} {\bibinfo
  {journal} {Phys. Rev. D}\ }\textbf {\bibinfo {volume} {99}},\ \bibinfo
  {pages} {064011} (\bibinfo {year} {2019})},\ \Eprint
  {https://arxiv.org/abs/1812.05590} {arXiv:1812.05590}\BibitemShut {NoStop}%
\bibitem [{\citenamefont {Macedo}\ \emph {et~al.}(2019)\citenamefont {Macedo},
  \citenamefont {Sakstein}, \citenamefont {Berti}, \citenamefont {Gualtieri},
  \citenamefont {Silva},\  and\ \citenamefont {Sotiriou}}]{Macedo:2019sem}%
  \BibitemOpen
  \bibfield  {author} {\bibinfo {author} {\bibfnamefont {C.~F.~B.}\
  \bibnamefont {Macedo}}, \bibinfo {author} {\bibfnamefont {J.}~\bibnamefont
  {Sakstein}}, \bibinfo {author} {\bibfnamefont {E.}~\bibnamefont {Berti}},
  \bibinfo {author} {\bibfnamefont {L.}~\bibnamefont {Gualtieri}}, \bibinfo
  {author} {\bibfnamefont {H.~O.}\ \bibnamefont {Silva}},  and\ \bibinfo
  {author} {\bibfnamefont {T.~P.}\ \bibnamefont {Sotiriou}},\ }\bibfield
  {title} {\bibinfo {title} {Self-interactions and spontaneous black hole
  scalarization},\ }\href {https://doi.org/10.1103/PhysRevD.99.104041}
  {\bibfield  {journal} {\bibinfo  {journal} {Phys. Rev. D}\ }\textbf {\bibinfo
  {volume} {99}},\ \bibinfo {pages} {104041} (\bibinfo {year} {2019})},\
  \Eprint {https://arxiv.org/abs/1903.06784} {arXiv:1903.06784}\BibitemShut
  {NoStop}%
\bibitem [{\citenamefont {Antoniou}\ \emph {et~al.}(2021)\citenamefont
  {Antoniou}, \citenamefont {Leh\'ebel}, \citenamefont {Ventagli},\  and\
  \citenamefont {Sotiriou}}]{Antoniou:2021zoy}%
  \BibitemOpen
  \bibfield  {author} {\bibinfo {author} {\bibfnamefont {G.}~\bibnamefont
  {Antoniou}}, \bibinfo {author} {\bibfnamefont {A.}~\bibnamefont {Leh\'ebel}},
  \bibinfo {author} {\bibfnamefont {G.}~\bibnamefont {Ventagli}},  and\
  \bibinfo {author} {\bibfnamefont {T.~P.}\ \bibnamefont {Sotiriou}},\
  }\bibfield  {title} {\bibinfo {title} {{Black hole scalarization with
  Gauss-Bonnet and Ricci scalar couplings}},\ }\href
  {https://doi.org/10.1103/PhysRevD.104.044002} {\bibfield  {journal} {\bibinfo
   {journal} {Phys. Rev. D}\ }\textbf {\bibinfo {volume} {104}},\ \bibinfo
  {pages} {044002} (\bibinfo {year} {2021})},\ \Eprint
  {https://arxiv.org/abs/2105.04479} {arXiv:2105.04479}\BibitemShut {NoStop}%
\bibitem [{\citenamefont {Herrero-Valea}(2022)}]{Herrero-Valea:2021dry}%
  \BibitemOpen
  \bibfield  {author} {\bibinfo {author} {\bibfnamefont {M.}~\bibnamefont
  {Herrero-Valea}},\ }\bibfield  {title} {\bibinfo {title} {{The shape of
  scalar Gauss-Bonnet gravity}},\ }\href
  {https://doi.org/10.1007/JHEP03(2022)075} {\bibfield  {journal} {\bibinfo
  {journal} {J. High Energy Phys.}\ }03 (2022)\ 075},\ \Eprint
  {https://arxiv.org/abs/2106.08344} {arXiv:2106.08344}\BibitemShut {NoStop}%
\bibitem [{\citenamefont {Herdeiro}\ \emph {et~al.}(2018)\citenamefont
  {Herdeiro}, \citenamefont {Radu}, \citenamefont {Sanchis-Gual},\  and\
  \citenamefont {Font}}]{Herdeiro:2018wub}%
  \BibitemOpen
  \bibfield  {author} {\bibinfo {author} {\bibfnamefont {C.~A.~R.}\
  \bibnamefont {Herdeiro}}, \bibinfo {author} {\bibfnamefont {E.}~\bibnamefont
  {Radu}}, \bibinfo {author} {\bibfnamefont {N.}~\bibnamefont {Sanchis-Gual}},
  and\ \bibinfo {author} {\bibfnamefont {J.~A.}\ \bibnamefont {Font}},\
  }\bibfield  {title} {\bibinfo {title} {{Spontaneous Scalarization of Charged
  Black Holes}},\ }\href {https://doi.org/10.1103/PhysRevLett.121.101102}
  {\bibfield  {journal} {\bibinfo  {journal} {Phys. Rev. Lett.}\ }\textbf
  {\bibinfo {volume} {121}},\ \bibinfo {pages} {101102} (\bibinfo {year}
  {2018})},\ \Eprint {https://arxiv.org/abs/1806.05190}
  {arXiv:1806.05190}\BibitemShut {NoStop}%
\bibitem [{\citenamefont {Andreou}\ \emph {et~al.}(2019)\citenamefont
  {Andreou}, \citenamefont {Franchini}, \citenamefont {Ventagli},\  and\
  \citenamefont {Sotiriou}}]{Andreou:2019ikc}%
  \BibitemOpen
  \bibfield  {author} {\bibinfo {author} {\bibfnamefont {N.}~\bibnamefont
  {Andreou}}, \bibinfo {author} {\bibfnamefont {N.}~\bibnamefont {Franchini}},
  \bibinfo {author} {\bibfnamefont {G.}~\bibnamefont {Ventagli}},  and\
  \bibinfo {author} {\bibfnamefont {T.~P.}\ \bibnamefont {Sotiriou}},\
  }\bibfield  {title} {\bibinfo {title} {{Spontaneous scalarization in
  generalised scalar-tensor theory}},\ }\href
  {https://doi.org/10.1103/PhysRevD.99.124022} {\bibfield  {journal} {\bibinfo
  {journal} {Phys. Rev. D}\ }\textbf {\bibinfo {volume} {99}},\ \bibinfo
  {pages} {124022} (\bibinfo {year} {2019})}\bibinfo {erratum}
  {{\href{https://link.aps.org/doi/10.1103/PhysRevD.101.109903}{; \textbf{101},
  109903(E) (2020)}}},\ \Eprint {https://arxiv.org/abs/1904.06365}
  {arXiv:1904.06365}\BibitemShut {NoStop}%
\bibitem [{\citenamefont {Astefanesei}\ \emph {et~al.}(2020)\citenamefont
  {Astefanesei}, \citenamefont {Herdeiro}, \citenamefont {Oliveira},\  and\
  \citenamefont {Radu}}]{Astefanesei:2020qxk}%
  \BibitemOpen
  \bibfield  {author} {\bibinfo {author} {\bibfnamefont {D.}~\bibnamefont
  {Astefanesei}}, \bibinfo {author} {\bibfnamefont {C.}~\bibnamefont
  {Herdeiro}}, \bibinfo {author} {\bibfnamefont {J.~a.}\ \bibnamefont
  {Oliveira}},  and\ \bibinfo {author} {\bibfnamefont {E.}~\bibnamefont
  {Radu}},\ }\bibfield  {title} {\bibinfo {title} {{Higher dimensional black
  hole scalarization}},\ }\href {https://doi.org/10.1007/JHEP09(2020)186}
  {\bibfield  {journal} {\bibinfo  {journal} {J. High Energy Phys.}\ }09
  (2020)\ 186},\ \Eprint {https://arxiv.org/abs/2007.04153}
  {arXiv:2007.04153}\BibitemShut {NoStop}%
\bibitem [{\citenamefont {Ventagli}\ \emph {et~al.}(2020)\citenamefont
  {Ventagli}, \citenamefont {Leh\'ebel},\  and\ \citenamefont
  {Sotiriou}}]{Ventagli:2020rnx}%
  \BibitemOpen
  \bibfield  {author} {\bibinfo {author} {\bibfnamefont {G.}~\bibnamefont
  {Ventagli}}, \bibinfo {author} {\bibfnamefont {A.}~\bibnamefont {Leh\'ebel}},
   and\ \bibinfo {author} {\bibfnamefont {T.~P.}\ \bibnamefont {Sotiriou}},\
  }\bibfield  {title} {\bibinfo {title} {{Onset of spontaneous scalarization in
  generalized scalar-tensor theories}},\ }\href
  {https://doi.org/10.1103/PhysRevD.102.024050} {\bibfield  {journal} {\bibinfo
   {journal} {Phys. Rev. D}\ }\textbf {\bibinfo {volume} {102}},\ \bibinfo
  {pages} {024050} (\bibinfo {year} {2020})},\ \Eprint
  {https://arxiv.org/abs/2006.01153} {arXiv:2006.01153}\BibitemShut {NoStop}%
\bibitem [{\citenamefont {Hod}(2019)}]{Hod:2019pmb}%
  \BibitemOpen
  \bibfield  {author} {\bibinfo {author} {\bibfnamefont {S.}~\bibnamefont
  {Hod}},\ }\bibfield  {title} {\bibinfo {title} {{Spontaneous scalarization of
  Gauss-Bonnet black holes: Analytic treatment in the linearized regime}},\
  }\href {https://doi.org/10.1103/PhysRevD.100.064039} {\bibfield  {journal}
  {\bibinfo  {journal} {Phys. Rev. D}\ }\textbf {\bibinfo {volume} {100}},\
  \bibinfo {pages} {064039} (\bibinfo {year} {2019})},\ \Eprint
  {https://arxiv.org/abs/1912.07630} {arXiv:1912.07630}\BibitemShut {NoStop}%
\bibitem [{\citenamefont {Herdeiro}\  and\ \citenamefont
  {Radu}(2019)}]{Herdeiro:2019yjy}%
  \BibitemOpen
  \bibfield  {author} {\bibinfo {author} {\bibfnamefont {C.~A.~R.}\
  \bibnamefont {Herdeiro}} and\ \bibinfo {author} {\bibfnamefont
  {E.}~\bibnamefont {Radu}},\ }\bibfield  {title} {\bibinfo {title} {{Black
  hole scalarization from the breakdown of scale invariance}},\ }\href
  {https://doi.org/10.1103/PhysRevD.99.084039} {\bibfield  {journal} {\bibinfo
  {journal} {Phys. Rev. D}\ }\textbf {\bibinfo {volume} {99}},\ \bibinfo
  {pages} {084039} (\bibinfo {year} {2019})},\ \Eprint
  {https://arxiv.org/abs/1901.02953} {arXiv:1901.02953}\BibitemShut {NoStop}%
\bibitem [{\citenamefont {Annulli}\ \emph {et~al.}(2022)\citenamefont
  {Annulli}, \citenamefont {Herdeiro},\  and\ \citenamefont
  {Radu}}]{Annulli:2022ivr}%
  \BibitemOpen
  \bibfield  {author} {\bibinfo {author} {\bibfnamefont {L.}~\bibnamefont
  {Annulli}}, \bibinfo {author} {\bibfnamefont {C.~A.~R.}\ \bibnamefont
  {Herdeiro}},  and\ \bibinfo {author} {\bibfnamefont {E.}~\bibnamefont
  {Radu}},\ }\bibfield  {title} {\bibinfo {title} {{Spin-induced scalarization
  and magnetic fields}},\ }\href
  {https://doi.org/10.1016/j.physletb.2022.137227} {\bibfield  {journal}
  {\bibinfo  {journal} {Phys. Lett. B}\ }\textbf {\bibinfo {volume} {832}},\
  \bibinfo {pages} {137227} (\bibinfo {year} {2022})},\ \Eprint
  {https://arxiv.org/abs/2203.13267} {arXiv:2203.13267}\BibitemShut {NoStop}%
\bibitem [{\citenamefont {Cherubini}\ \emph {et~al.}(2002)\citenamefont
  {Cherubini}, \citenamefont {Bini}, \citenamefont {Capozziello},\  and\
  \citenamefont {Ruffini}}]{Cherubini:2002gen}%
  \BibitemOpen
  \bibfield  {author} {\bibinfo {author} {\bibfnamefont {C.}~\bibnamefont
  {Cherubini}}, \bibinfo {author} {\bibfnamefont {D.}~\bibnamefont {Bini}},
  \bibinfo {author} {\bibfnamefont {S.}~\bibnamefont {Capozziello}},  and\
  \bibinfo {author} {\bibfnamefont {R.}~\bibnamefont {Ruffini}},\ }\bibfield
  {title} {\bibinfo {title} {{Second order scalar invariants of the Riemann
  tensor: Applications to black hole spacetimes}},\ }\href
  {https://doi.org/10.1142/S0218271802002037} {\bibfield  {journal} {\bibinfo
  {journal} {Int. J. Mod. Phys. D}\ }\textbf {\bibinfo {volume} {11}},\
  \bibinfo {pages} {827} (\bibinfo {year} {2002})},\ \Eprint
  {https://arxiv.org/abs/gr-qc/0302095} {arXiv:gr-qc/0302095}\BibitemShut
  {NoStop}%
\bibitem [{\citenamefont {Abbott}\ \emph
  {et~al.}(2016{\natexlab{b}})\citenamefont {Abbott} \emph
  {et~al.}}]{TheLIGOScientific:2016src}%
  \BibitemOpen
  \bibfield  {author} {\bibinfo {author} {\bibfnamefont {B.~P.}\ \bibnamefont
  {Abbott}} \emph {et~al.} (\bibinfo {collaboration} {LIGO Scientific and Virgo
  Collaborations}),\ }\bibfield  {title} {\bibinfo {title} {{Tests of General
  Relativity with GW150914}},\ }\href
  {https://doi.org/10.1103/PhysRevLett.116.221101} {\bibfield  {journal}
  {\bibinfo  {journal} {Phys. Rev. Lett.}\ }\textbf {\bibinfo {volume} {116}},\
  \bibinfo {pages} {221101} (\bibinfo {year} {2016}{\natexlab{b}})}\bibinfo
  {erratum} {{\href{https://link.aps.org/doi/10.1103/PhysRevLett.121.129902}{;
  \textbf{121}, 129902(E) (2018)}}},\ \Eprint
  {https://arxiv.org/abs/1602.03841} {arXiv:1602.03841}\BibitemShut {NoStop}%
\bibitem [{\citenamefont {Abbott}\ \emph
  {et~al.}(2019{\natexlab{a}})\citenamefont {Abbott} \emph
  {et~al.}}]{LIGOScientific:2018dkp}%
  \BibitemOpen
  \bibfield  {author} {\bibinfo {author} {\bibfnamefont {B.~P.}\ \bibnamefont
  {Abbott}} \emph {et~al.} (\bibinfo {collaboration} {LIGO Scientific and Virgo
  Collaborations}),\ }\bibfield  {title} {\bibinfo {title} {{Tests of General
  Relativity with GW170817}},\ }\href
  {https://doi.org/10.1103/PhysRevLett.123.011102} {\bibfield  {journal}
  {\bibinfo  {journal} {Phys. Rev. Lett.}\ }\textbf {\bibinfo {volume} {123}},\
  \bibinfo {pages} {011102} (\bibinfo {year} {2019}{\natexlab{a}})},\ \Eprint
  {https://arxiv.org/abs/1811.00364} {arXiv:1811.00364}\BibitemShut {NoStop}%
\bibitem [{\citenamefont {Abbott}\ \emph
  {et~al.}(2019{\natexlab{b}})\citenamefont {Abbott} \emph
  {et~al.}}]{LIGOScientific:2019fpa}%
  \BibitemOpen
  \bibfield  {author} {\bibinfo {author} {\bibfnamefont {B.~P.}\ \bibnamefont
  {Abbott}} \emph {et~al.} (\bibinfo {collaboration} {LIGO Scientific and Virgo
  Collaborations}),\ }\bibfield  {title} {\bibinfo {title} {Tests of general
  relativity with the binary black hole signals from the {LIGO-Virgo catalog
  GWTC-1}},\ }\href {https://doi.org/10.1103/PhysRevD.100.104036} {\bibfield
  {journal} {\bibinfo  {journal} {Phys. Rev. D}\ }\textbf {\bibinfo {volume}
  {100}},\ \bibinfo {pages} {104036} (\bibinfo {year} {2019}{\natexlab{b}})},\
  \Eprint {https://arxiv.org/abs/1903.04467} {arXiv:1903.04467}\BibitemShut
  {NoStop}%
\bibitem [{\citenamefont {Abbott}\ \emph
  {et~al.}(2021{\natexlab{a}})\citenamefont {Abbott} \emph
  {et~al.}}]{LIGOScientific:2020tif}%
  \BibitemOpen
  \bibfield  {author} {\bibinfo {author} {\bibfnamefont {R.}~\bibnamefont
  {Abbott}} \emph {et~al.} (\bibinfo {collaboration} {LIGO Scientific and Virgo
  Collaborations}),\ }\bibfield  {title} {\bibinfo {title} {Tests of general
  relativity with binary black holes from the second {LIGO-Virgo
  gravitational-wave transient catalog}},\ }\href
  {https://doi.org/10.1103/PhysRevD.103.122002} {\bibfield  {journal} {\bibinfo
   {journal} {Phys. Rev. D}\ }\textbf {\bibinfo {volume} {103}},\ \bibinfo
  {pages} {122002} (\bibinfo {year} {2021}{\natexlab{a}})},\ \Eprint
  {https://arxiv.org/abs/2010.14529} {arXiv:2010.14529}\BibitemShut {NoStop}%
\bibitem [{\citenamefont {Abbott}\ \emph
  {et~al.}(2021{\natexlab{b}})\citenamefont {Abbott} \emph
  {et~al.}}]{LIGOScientific:2021sio}%
  \BibitemOpen
  \bibfield  {author} {\bibinfo {author} {\bibfnamefont {R.}~\bibnamefont
  {Abbott}} \emph {et~al.} (\bibinfo {collaboration} {LIGO Scientific, Virgo,
  and KAGRA Collaborations}),\ }\bibfield  {title} {\bibinfo {title} {{Tests of
  general relativity with GWTC-3}},\ }\Eprint
  {https://arxiv.org/abs/2112.06861} {arXiv:2112.06861}\BibitemShut {NoStop}%
\bibitem [{\citenamefont {Abbott}\ \emph
  {et~al.}(2020{\natexlab{b}})\citenamefont {Abbott} \emph
  {et~al.}}]{LIGOScientific:2019hgc}%
  \BibitemOpen
  \bibfield  {author} {\bibinfo {author} {\bibfnamefont {B.~P.}\ \bibnamefont
  {Abbott}} \emph {et~al.} (\bibinfo {collaboration} {LIGO Scientific and Virgo
  Collaborations}),\ }\bibfield  {title} {\bibinfo {title} {{A guide to
  LIGO\textendash{}Virgo detector noise and extraction of transient
  gravitational-wave signals}},\ }\href
  {https://doi.org/10.1088/1361-6382/ab685e} {\bibfield  {journal} {\bibinfo
  {journal} {Classical Quantum Gravity}\ }\textbf {\bibinfo {volume} {37}},\
  \bibinfo {pages} {055002} (\bibinfo {year} {2020}{\natexlab{b}})},\ \Eprint
  {https://arxiv.org/abs/1908.11170} {arXiv:1908.11170}\BibitemShut {NoStop}%
\bibitem [{\citenamefont {Cutler}\  and\ \citenamefont
  {Flanagan}(1994)}]{Cutler:1994ys}%
  \BibitemOpen
  \bibfield  {author} {\bibinfo {author} {\bibfnamefont {C.}~\bibnamefont
  {Cutler}} and\ \bibinfo {author} {\bibfnamefont {E.~E.}\ \bibnamefont
  {Flanagan}},\ }\bibfield  {title} {\bibinfo {title} {Gravitational waves from
  merging compact binaries: How accurately can one extract the binary's
  parameters from the inspiral waveform?},\ }\href
  {https://doi.org/10.1103/PhysRevD.49.2658} {\bibfield  {journal} {\bibinfo
  {journal} {Phys. Rev. D}\ }\textbf {\bibinfo {volume} {49}},\ \bibinfo
  {pages} {2658} (\bibinfo {year} {1994})},\ \Eprint
  {https://arxiv.org/abs/gr-qc/9402014} {arXiv:gr-qc/9402014}\BibitemShut
  {NoStop}%
\bibitem [{\citenamefont {Veitch}\ \emph {et~al.}(2015)\citenamefont {Veitch},
  \citenamefont {Raymond}, \citenamefont {Farr}, \citenamefont {Farr},
  \citenamefont {Graff}, \citenamefont {Vitale}, \citenamefont {Aylott},
  \citenamefont {Blackburn}, \citenamefont {Christensen}, \citenamefont
  {Coughlin} \emph {et~al.}}]{Veitch:2014wba}%
  \BibitemOpen
  \bibfield  {author} {\bibinfo {author} {\bibfnamefont {J.}~\bibnamefont
  {Veitch}}, \bibinfo {author} {\bibfnamefont {V.}~\bibnamefont {Raymond}},
  \bibinfo {author} {\bibfnamefont {B.}~\bibnamefont {Farr}}, \bibinfo {author}
  {\bibfnamefont {W.}~\bibnamefont {Farr}}, \bibinfo {author} {\bibfnamefont
  {P.}~\bibnamefont {Graff}}, \bibinfo {author} {\bibfnamefont
  {S.}~\bibnamefont {Vitale}}, \bibinfo {author} {\bibfnamefont
  {B.}~\bibnamefont {Aylott}}, \bibinfo {author} {\bibfnamefont
  {K.}~\bibnamefont {Blackburn}}, \bibinfo {author} {\bibfnamefont
  {N.}~\bibnamefont {Christensen}}, \bibinfo {author} {\bibfnamefont
  {M.}~\bibnamefont {Coughlin}},  \emph {et~al.},\ }\bibfield  {title}
  {\bibinfo {title} {{Parameter estimation for compact binaries with
  ground-based gravitational-wave observations using the LALInference software
  library}},\ }\href {https://doi.org/10.1103/PhysRevD.91.042003} {\bibfield
  {journal} {\bibinfo  {journal} {Phys. Rev. D}\ }\textbf {\bibinfo {volume}
  {91}},\ \bibinfo {pages} {042003} (\bibinfo {year} {2015})},\ \Eprint
  {https://arxiv.org/abs/1409.7215} {arXiv:1409.7215}\BibitemShut {NoStop}%
\bibitem [{\citenamefont {Yunes}\  and\ \citenamefont
  {Pretorius}(2009)}]{Yunes:2009ke}%
  \BibitemOpen
  \bibfield  {author} {\bibinfo {author} {\bibfnamefont {N.}~\bibnamefont
  {Yunes}} and\ \bibinfo {author} {\bibfnamefont {F.}~\bibnamefont
  {Pretorius}},\ }\bibfield  {title} {\bibinfo {title} {Fundamental theoretical
  bias in gravitational wave astrophysics and the parameterized
  post-{E}insteinian framework},\ }\href
  {https://doi.org/10.1103/PhysRevD.80.122003} {\bibfield  {journal} {\bibinfo
  {journal} {Phys. Rev. D}\ }\textbf {\bibinfo {volume} {80}},\ \bibinfo
  {pages} {122003} (\bibinfo {year} {2009})},\ \Eprint
  {https://arxiv.org/abs/0909.3328} {arXiv:0909.3328}\BibitemShut {NoStop}%
\bibitem [{\citenamefont {Damour}\  and\ \citenamefont
  {Esposito-Far{\`e}se}(1992)}]{Damour:1992we}%
  \BibitemOpen
  \bibfield  {author} {\bibinfo {author} {\bibfnamefont {T.}~\bibnamefont
  {Damour}} and\ \bibinfo {author} {\bibfnamefont {G.}~\bibnamefont
  {Esposito-Far{\`e}se}},\ }\bibfield  {title} {\bibinfo {title}
  {{Tensor-multi-scalar theories of gravitation}},\ }\href
  {https://doi.org/10.1088/0264-9381/9/9/015} {\bibfield  {journal} {\bibinfo
  {journal} {Classical Quantum Gravity}\ }\textbf {\bibinfo {volume} {9}},\
  \bibinfo {pages} {2093} (\bibinfo {year} {1992})}\BibitemShut {NoStop}%
\bibitem [{\citenamefont {Mirshekari}\  and\ \citenamefont
  {Will}(2013)}]{Mirshekari:2013vb}%
  \BibitemOpen
  \bibfield  {author} {\bibinfo {author} {\bibfnamefont {S.}~\bibnamefont
  {Mirshekari}} and\ \bibinfo {author} {\bibfnamefont {C.~M.}\ \bibnamefont
  {Will}},\ }\bibfield  {title} {\bibinfo {title} {{Compact binary systems in
  scalar-tensor gravity: Equations of motion to 2.5 post-Newtonian order}},\
  }\href {https://doi.org/10.1103/PhysRevD.87.084070} {\bibfield  {journal}
  {\bibinfo  {journal} {Phys. Rev. D}\ }\textbf {\bibinfo {volume} {87}},\
  \bibinfo {pages} {084070} (\bibinfo {year} {2013})},\ \Eprint
  {https://arxiv.org/abs/1301.4680} {arXiv:1301.4680}\BibitemShut {NoStop}%
\bibitem [{\citenamefont {Lang}(2014)}]{Lang:2013fna}%
  \BibitemOpen
  \bibfield  {author} {\bibinfo {author} {\bibfnamefont {R.~N.}\ \bibnamefont
  {Lang}},\ }\bibfield  {title} {\bibinfo {title} {{Compact binary systems in
  scalar-tensor gravity. II. Tensor gravitational waves to second
  post-Newtonian order}},\ }\href {https://doi.org/10.1103/PhysRevD.89.084014}
  {\bibfield  {journal} {\bibinfo  {journal} {Phys. Rev. D}\ }\textbf {\bibinfo
  {volume} {89}},\ \bibinfo {pages} {084014} (\bibinfo {year} {2014})},\
  \Eprint {https://arxiv.org/abs/1310.3320} {arXiv:1310.3320}\BibitemShut
  {NoStop}%
\bibitem [{\citenamefont {Lang}(2015)}]{Lang:2014osa}%
  \BibitemOpen
  \bibfield  {author} {\bibinfo {author} {\bibfnamefont {R.~N.}\ \bibnamefont
  {Lang}},\ }\bibfield  {title} {\bibinfo {title} {{Compact binary systems in
  scalar-tensor gravity. III. Scalar waves and energy flux}},\ }\href
  {https://doi.org/10.1103/PhysRevD.91.084027} {\bibfield  {journal} {\bibinfo
  {journal} {Phys. Rev. D}\ }\textbf {\bibinfo {volume} {91}},\ \bibinfo
  {pages} {084027} (\bibinfo {year} {2015})},\ \Eprint
  {https://arxiv.org/abs/1411.3073} {arXiv:1411.3073}\BibitemShut {NoStop}%
\bibitem [{\citenamefont {Bernard}(2018)}]{Bernard:2018hta}%
  \BibitemOpen
  \bibfield  {author} {\bibinfo {author} {\bibfnamefont {L.}~\bibnamefont
  {Bernard}},\ }\bibfield  {title} {\bibinfo {title} {{Dynamics of compact
  binary systems in scalar-tensor theories: Equations of motion to the third
  post-Newtonian order}},\ }\href {https://doi.org/10.1103/PhysRevD.98.044004}
  {\bibfield  {journal} {\bibinfo  {journal} {Phys. Rev. D}\ }\textbf {\bibinfo
  {volume} {98}},\ \bibinfo {pages} {044004} (\bibinfo {year} {2018})},\
  \Eprint {https://arxiv.org/abs/1802.10201} {arXiv:1802.10201}\BibitemShut
  {NoStop}%
\bibitem [{\citenamefont {Bernard}(2019)}]{Bernard:2018ivi}%
  \BibitemOpen
  \bibfield  {author} {\bibinfo {author} {\bibfnamefont {L.}~\bibnamefont
  {Bernard}},\ }\bibfield  {title} {\bibinfo {title} {{Dynamics of compact
  binary systems in scalar-tensor theories: II. Center-of-mass and conserved
  quantities to 3PN order}},\ }\href
  {https://doi.org/10.1103/PhysRevD.99.044047} {\bibfield  {journal} {\bibinfo
  {journal} {Phys. Rev. D}\ }\textbf {\bibinfo {volume} {99}},\ \bibinfo
  {pages} {044047} (\bibinfo {year} {2019})},\ \Eprint
  {https://arxiv.org/abs/1812.04169} {arXiv:1812.04169}\BibitemShut {NoStop}%
\bibitem [{\citenamefont {Sennett}\ \emph {et~al.}(2016)\citenamefont
  {Sennett}, \citenamefont {Marsat},\  and\ \citenamefont
  {Buonanno}}]{Sennett:2016klh}%
  \BibitemOpen
  \bibfield  {author} {\bibinfo {author} {\bibfnamefont {N.}~\bibnamefont
  {Sennett}}, \bibinfo {author} {\bibfnamefont {S.}~\bibnamefont {Marsat}},
  and\ \bibinfo {author} {\bibfnamefont {A.}~\bibnamefont {Buonanno}},\
  }\bibfield  {title} {\bibinfo {title} {{Gravitational waveforms in
  scalar-tensor gravity at 2PN relative order}},\ }\href
  {https://doi.org/10.1103/PhysRevD.94.084003} {\bibfield  {journal} {\bibinfo
  {journal} {Phys. Rev. D}\ }\textbf {\bibinfo {volume} {94}},\ \bibinfo
  {pages} {084003} (\bibinfo {year} {2016})},\ \Eprint
  {https://arxiv.org/abs/1607.01420} {arXiv:1607.01420}\BibitemShut {NoStop}%
\bibitem [{\citenamefont {Shao}\ \emph {et~al.}(2017)\citenamefont {Shao},
  \citenamefont {Sennett}, \citenamefont {Buonanno}, \citenamefont {Kramer},\
  and\ \citenamefont {Wex}}]{Shao:2017gwu}%
  \BibitemOpen
  \bibfield  {author} {\bibinfo {author} {\bibfnamefont {L.}~\bibnamefont
  {Shao}}, \bibinfo {author} {\bibfnamefont {N.}~\bibnamefont {Sennett}},
  \bibinfo {author} {\bibfnamefont {A.}~\bibnamefont {Buonanno}}, \bibinfo
  {author} {\bibfnamefont {M.}~\bibnamefont {Kramer}},  and\ \bibinfo {author}
  {\bibfnamefont {N.}~\bibnamefont {Wex}},\ }\bibfield  {title} {\bibinfo
  {title} {{Constraining Nonperturbative Strong-Field Effects in Scalar-Tensor
  Gravity by Combining Pulsar Timing and Laser-Interferometer
  Gravitational-Wave Detectors}},\ }\href
  {https://doi.org/10.1103/PhysRevX.7.041025} {\bibfield  {journal} {\bibinfo
  {journal} {Phys. Rev. X}\ }\textbf {\bibinfo {volume} {7}},\ \bibinfo {pages}
  {041025} (\bibinfo {year} {2017})},\ \Eprint
  {https://arxiv.org/abs/1704.07561} {arXiv:1704.07561}\BibitemShut {NoStop}%
\bibitem [{\citenamefont {Huang}\ \emph {et~al.}(2019)\citenamefont {Huang},
  \citenamefont {Johnson}, \citenamefont {Sagunski}, \citenamefont
  {Sakellariadou},\  and\ \citenamefont {Zhang}}]{Huang:2018pbu}%
  \BibitemOpen
  \bibfield  {author} {\bibinfo {author} {\bibfnamefont {J.}~\bibnamefont
  {Huang}}, \bibinfo {author} {\bibfnamefont {M.~C.}\ \bibnamefont {Johnson}},
  \bibinfo {author} {\bibfnamefont {L.}~\bibnamefont {Sagunski}}, \bibinfo
  {author} {\bibfnamefont {M.}~\bibnamefont {Sakellariadou}},  and\ \bibinfo
  {author} {\bibfnamefont {J.}~\bibnamefont {Zhang}},\ }\bibfield  {title}
  {\bibinfo {title} {{Prospects for axion searches with Advanced LIGO through
  binary mergers}},\ }\href {https://doi.org/10.1103/PhysRevD.99.063013}
  {\bibfield  {journal} {\bibinfo  {journal} {Phys. Rev. D}\ }\textbf {\bibinfo
  {volume} {99}},\ \bibinfo {pages} {063013} (\bibinfo {year} {2019})},\
  \Eprint {https://arxiv.org/abs/1807.02133} {arXiv:1807.02133}\BibitemShut
  {NoStop}%
\bibitem [{\citenamefont {Kuntz}\ \emph {et~al.}(2019)\citenamefont {Kuntz},
  \citenamefont {Piazza},\  and\ \citenamefont {Vernizzi}}]{Kuntz:2019zef}%
  \BibitemOpen
  \bibfield  {author} {\bibinfo {author} {\bibfnamefont {A.}~\bibnamefont
  {Kuntz}}, \bibinfo {author} {\bibfnamefont {F.}~\bibnamefont {Piazza}},  and\
  \bibinfo {author} {\bibfnamefont {F.}~\bibnamefont {Vernizzi}},\ }\bibfield
  {title} {\bibinfo {title} {{Effective field theory for gravitational
  radiation in scalar-tensor gravity}},\ }\href
  {https://doi.org/10.1088/1475-7516/2019/05/052} {\bibfield  {journal}
  {\bibinfo  {journal} {J. Cosmol. Astropart. Phys.}\ }05 (2019)\ 052},\
  \Eprint {https://arxiv.org/abs/1902.04941} {arXiv:1902.04941}\BibitemShut
  {NoStop}%
\bibitem [{\citenamefont {Brax}\ \emph {et~al.}(2021)\citenamefont {Brax},
  \citenamefont {Davis}, \citenamefont {Melville},\  and\ \citenamefont
  {Wong}}]{Brax:2021qqo}%
  \BibitemOpen
  \bibfield  {author} {\bibinfo {author} {\bibfnamefont {P.}~\bibnamefont
  {Brax}}, \bibinfo {author} {\bibfnamefont {A.-C.}\ \bibnamefont {Davis}},
  \bibinfo {author} {\bibfnamefont {S.}~\bibnamefont {Melville}},  and\
  \bibinfo {author} {\bibfnamefont {L.~K.}\ \bibnamefont {Wong}},\ }\bibfield
  {title} {\bibinfo {title} {{Spin-orbit effects for compact binaries in
  scalar-tensor gravity}},\ }\href
  {https://doi.org/10.1088/1475-7516/2021/10/075} {\bibfield  {journal}
  {\bibinfo  {journal} {J. Cosmol. Astropart. Phys.}\ }10 (2021)\ 075},\
  \Eprint {https://arxiv.org/abs/2107.10841} {arXiv:2107.10841}\BibitemShut
  {NoStop}%
\bibitem [{\citenamefont {Bernard}\ \emph {et~al.}(2022)\citenamefont
  {Bernard}, \citenamefont {Blanchet},\  and\ \citenamefont
  {Trestini}}]{Bernard:2022noq}%
  \BibitemOpen
  \bibfield  {author} {\bibinfo {author} {\bibfnamefont {L.}~\bibnamefont
  {Bernard}}, \bibinfo {author} {\bibfnamefont {L.}~\bibnamefont {Blanchet}},
  and\ \bibinfo {author} {\bibfnamefont {D.}~\bibnamefont {Trestini}},\
  }\bibfield  {title} {\bibinfo {title} {{Gravitational waves in scalar-tensor
  theory to one-and-a-half post-Newtonian order}},\ }\Eprint
  {https://arxiv.org/abs/2201.10924} {arXiv:2201.10924}\BibitemShut {NoStop}%
\bibitem [{\citenamefont {Yagi}\ \emph {et~al.}(2012)\citenamefont {Yagi},
  \citenamefont {Stein}, \citenamefont {Yunes},\  and\ \citenamefont
  {Tanaka}}]{Yagi:2011xp}%
  \BibitemOpen
  \bibfield  {author} {\bibinfo {author} {\bibfnamefont {K.}~\bibnamefont
  {Yagi}}, \bibinfo {author} {\bibfnamefont {L.~C.}\ \bibnamefont {Stein}},
  \bibinfo {author} {\bibfnamefont {N.}~\bibnamefont {Yunes}},  and\ \bibinfo
  {author} {\bibfnamefont {T.}~\bibnamefont {Tanaka}},\ }\bibfield  {title}
  {\bibinfo {title} {{Post-Newtonian, quasicircular binary inspirals in
  quadratic modified gravity}},\ }\href
  {https://doi.org/10.1103/PhysRevD.85.064022} {\bibfield  {journal} {\bibinfo
  {journal} {Phys. Rev. D}\ }\textbf {\bibinfo {volume} {85}},\ \bibinfo
  {pages} {064022} (\bibinfo {year} {2012})}\bibinfo {erratum}
  {{\href{https://link.aps.org/doi/10.1103/PhysRevD.93.029902}{; \textbf{93},
  029902 (2016)}}},\ \Eprint {https://arxiv.org/abs/1110.5950}
  {arXiv:1110.5950}\BibitemShut {NoStop}%
\bibitem [{\citenamefont {Juli\'e}\  and\ \citenamefont
  {Berti}(2019)}]{Julie:2019sab}%
  \BibitemOpen
  \bibfield  {author} {\bibinfo {author} {\bibfnamefont {F.-L.}\ \bibnamefont
  {Juli\'e}} and\ \bibinfo {author} {\bibfnamefont {E.}~\bibnamefont {Berti}},\
  }\bibfield  {title} {\bibinfo {title} {{Post-Newtonian dynamics and black
  hole thermodynamics in Einstein-scalar-Gauss-Bonnet gravity}},\ }\href
  {https://doi.org/10.1103/PhysRevD.100.104061} {\bibfield  {journal} {\bibinfo
   {journal} {Phys. Rev. D}\ }\textbf {\bibinfo {volume} {100}},\ \bibinfo
  {pages} {104061} (\bibinfo {year} {2019})},\ \Eprint
  {https://arxiv.org/abs/1909.05258} {arXiv:1909.05258}\BibitemShut {NoStop}%
\bibitem [{\citenamefont {Shiralilou}\ \emph {et~al.}(2021)\citenamefont
  {Shiralilou}, \citenamefont {Hinderer}, \citenamefont {Nissanke},
  \citenamefont {Ortiz},\  and\ \citenamefont {Witek}}]{Shiralilou:2020gah}%
  \BibitemOpen
  \bibfield  {author} {\bibinfo {author} {\bibfnamefont {B.}~\bibnamefont
  {Shiralilou}}, \bibinfo {author} {\bibfnamefont {T.}~\bibnamefont
  {Hinderer}}, \bibinfo {author} {\bibfnamefont {S.~M.}\ \bibnamefont
  {Nissanke}}, \bibinfo {author} {\bibfnamefont {N.}~\bibnamefont {Ortiz}},
  and\ \bibinfo {author} {\bibfnamefont {H.}~\bibnamefont {Witek}},\ }\bibfield
   {title} {\bibinfo {title} {{Nonlinear curvature effects in gravitational
  waves from inspiralling black hole binaries}},\ }\href
  {https://doi.org/10.1103/PhysRevD.103.L121503} {\bibfield  {journal}
  {\bibinfo  {journal} {Phys. Rev. D}\ }\textbf {\bibinfo {volume} {103}},\
  \bibinfo {pages} {L121503} (\bibinfo {year} {2021})},\ \Eprint
  {https://arxiv.org/abs/2012.09162} {arXiv:2012.09162}\BibitemShut {NoStop}%
\bibitem [{\citenamefont {Shiralilou}\ \emph {et~al.}(2022)\citenamefont
  {Shiralilou}, \citenamefont {Hinderer}, \citenamefont {Nissanke},
  \citenamefont {Ortiz},\  and\ \citenamefont {Witek}}]{Shiralilou:2021mfl}%
  \BibitemOpen
  \bibfield  {author} {\bibinfo {author} {\bibfnamefont {B.}~\bibnamefont
  {Shiralilou}}, \bibinfo {author} {\bibfnamefont {T.}~\bibnamefont
  {Hinderer}}, \bibinfo {author} {\bibfnamefont {S.~M.}\ \bibnamefont
  {Nissanke}}, \bibinfo {author} {\bibfnamefont {N.}~\bibnamefont {Ortiz}},
  and\ \bibinfo {author} {\bibfnamefont {H.}~\bibnamefont {Witek}},\ }\bibfield
   {title} {\bibinfo {title} {{Post-Newtonian gravitational and scalar waves in
  scalar-Gauss\textendash{}Bonnet gravity}},\ }\href
  {https://doi.org/10.1088/1361-6382/ac4196} {\bibfield  {journal} {\bibinfo
  {journal} {Classical Quantum Gravity}\ }\textbf {\bibinfo {volume} {39}},\
  \bibinfo {pages} {035002} (\bibinfo {year} {2022})},\ \Eprint
  {https://arxiv.org/abs/2105.13972} {arXiv:2105.13972}\BibitemShut {NoStop}%
\bibitem [{\citenamefont {Perkins}\  and\ \citenamefont
  {Yunes}(2022)}]{Perkins:2022fhr}%
  \BibitemOpen
  \bibfield  {author} {\bibinfo {author} {\bibfnamefont {S.}~\bibnamefont
  {Perkins}} and\ \bibinfo {author} {\bibfnamefont {N.}~\bibnamefont {Yunes}},\
  }\bibfield  {title} {\bibinfo {title} {Are parametrized tests of general
  relativity with gravitational waves robust to unknown higher post-{N}ewtonian
  order effects?},\ }\Eprint {https://arxiv.org/abs/2201.02542}
  {arXiv:2201.02542}\BibitemShut {NoStop}%
\bibitem [{\citenamefont {Tahura}\ \emph {et~al.}(2019)\citenamefont {Tahura},
  \citenamefont {Yagi},\  and\ \citenamefont {Carson}}]{Tahura:2019dgr}%
  \BibitemOpen
  \bibfield  {author} {\bibinfo {author} {\bibfnamefont {S.}~\bibnamefont
  {Tahura}}, \bibinfo {author} {\bibfnamefont {K.}~\bibnamefont {Yagi}},  and\
  \bibinfo {author} {\bibfnamefont {Z.}~\bibnamefont {Carson}},\ }\bibfield
  {title} {\bibinfo {title} {{Testing gravity with gravitational waves from
  binary black hole mergers: Contributions from amplitude corrections}},\
  }\href {https://doi.org/10.1103/PhysRevD.100.104001} {\bibfield  {journal}
  {\bibinfo  {journal} {Phys. Rev. D}\ }\textbf {\bibinfo {volume} {100}},\
  \bibinfo {pages} {104001} (\bibinfo {year} {2019})},\ \Eprint
  {https://arxiv.org/abs/1907.10059} {arXiv:1907.10059}\BibitemShut {NoStop}%
\bibitem [{\citenamefont {Sennett}\ \emph {et~al.}(2017)\citenamefont
  {Sennett}, \citenamefont {Shao},\  and\ \citenamefont
  {Steinhoff}}]{Sennett:2017lcx}%
  \BibitemOpen
  \bibfield  {author} {\bibinfo {author} {\bibfnamefont {N.}~\bibnamefont
  {Sennett}}, \bibinfo {author} {\bibfnamefont {L.}~\bibnamefont {Shao}},  and\
  \bibinfo {author} {\bibfnamefont {J.}~\bibnamefont {Steinhoff}},\ }\bibfield
  {title} {\bibinfo {title} {Effective action model of dynamically scalarizing
  binary neutron stars},\ }\href {https://doi.org/10.1103/PhysRevD.96.084019}
  {\bibfield  {journal} {\bibinfo  {journal} {Phys. Rev. D}\ }\textbf {\bibinfo
  {volume} {96}},\ \bibinfo {pages} {084019} (\bibinfo {year} {2017})},\
  \Eprint {https://arxiv.org/abs/1708.08285} {arXiv:1708.08285}\BibitemShut
  {NoStop}%
\bibitem [{\citenamefont {Juli\'e}\ \emph {et~al.}(2022)\citenamefont
  {Juli\'e}, \citenamefont {Silva}, \citenamefont {Berti},\  and\ \citenamefont
  {Yunes}}]{Julie:2022huo}%
  \BibitemOpen
  \bibfield  {author} {\bibinfo {author} {\bibfnamefont {F.-L.}\ \bibnamefont
  {Juli\'e}}, \bibinfo {author} {\bibfnamefont {H.~O.}\ \bibnamefont {Silva}},
  \bibinfo {author} {\bibfnamefont {E.}~\bibnamefont {Berti}},  and\ \bibinfo
  {author} {\bibfnamefont {N.}~\bibnamefont {Yunes}},\ }\bibfield  {title}
  {\bibinfo {title} {{Black hole sensitivities in Einstein-scalar-Gauss-Bonnet
  gravity}},\ }\href {https://doi.org/10.1103/PhysRevD.105.124031} {\bibfield
  {journal} {\bibinfo  {journal} {Phys. Rev. D}\ }\textbf {\bibinfo {volume}
  {105}},\ \bibinfo {pages} {124031} (\bibinfo {year} {2022})},\ \Eprint
  {https://arxiv.org/abs/2202.01329} {arXiv:2202.01329}\BibitemShut {NoStop}%
\bibitem [{\citenamefont {Carson}\  and\ \citenamefont
  {Yagi}(2020)}]{Carson:2020ter}%
  \BibitemOpen
  \bibfield  {author} {\bibinfo {author} {\bibfnamefont {Z.}~\bibnamefont
  {Carson}} and\ \bibinfo {author} {\bibfnamefont {K.}~\bibnamefont {Yagi}},\
  }\bibfield  {title} {\bibinfo {title} {{Probing Einstein-dilaton Gauss-Bonnet
  gravity with the inspiral and ringdown of gravitational waves}},\ }\href
  {https://doi.org/10.1103/PhysRevD.101.104030} {\bibfield  {journal} {\bibinfo
   {journal} {Phys. Rev. D}\ }\textbf {\bibinfo {volume} {101}},\ \bibinfo
  {pages} {104030} (\bibinfo {year} {2020})},\ \Eprint
  {https://arxiv.org/abs/2003.00286} {arXiv:2003.00286}\BibitemShut {NoStop}%
\bibitem [{\citenamefont {Witek}\ \emph {et~al.}(2019)\citenamefont {Witek},
  \citenamefont {Gualtieri}, \citenamefont {Pani},\  and\ \citenamefont
  {Sotiriou}}]{Witek:2018dmd}%
  \BibitemOpen
  \bibfield  {author} {\bibinfo {author} {\bibfnamefont {H.}~\bibnamefont
  {Witek}}, \bibinfo {author} {\bibfnamefont {L.}~\bibnamefont {Gualtieri}},
  \bibinfo {author} {\bibfnamefont {P.}~\bibnamefont {Pani}},  and\ \bibinfo
  {author} {\bibfnamefont {T.~P.}\ \bibnamefont {Sotiriou}},\ }\bibfield
  {title} {\bibinfo {title} {{Black holes and binary mergers in scalar
  Gauss-Bonnet gravity: Scalar field dynamics}},\ }\href
  {https://doi.org/10.1103/PhysRevD.99.064035} {\bibfield  {journal} {\bibinfo
  {journal} {Phys. Rev. D}\ }\textbf {\bibinfo {volume} {99}},\ \bibinfo
  {pages} {064035} (\bibinfo {year} {2019})},\ \Eprint
  {https://arxiv.org/abs/1810.05177} {arXiv:1810.05177}\BibitemShut {NoStop}%
\bibitem [{\citenamefont {Witek}\ \emph {et~al.}(2020)\citenamefont {Witek},
  \citenamefont {Gualtieri},\  and\ \citenamefont {Pani}}]{Witek:2020uzz}%
  \BibitemOpen
  \bibfield  {author} {\bibinfo {author} {\bibfnamefont {H.}~\bibnamefont
  {Witek}}, \bibinfo {author} {\bibfnamefont {L.}~\bibnamefont {Gualtieri}},
  and\ \bibinfo {author} {\bibfnamefont {P.}~\bibnamefont {Pani}},\ }\bibfield
  {title} {\bibinfo {title} {{Towards numerical relativity in scalar
  Gauss-Bonnet gravity: $3+1$ decomposition beyond the small-coupling limit}},\
  }\href {https://doi.org/10.1103/PhysRevD.101.124055} {\bibfield  {journal}
  {\bibinfo  {journal} {Phys. Rev. D}\ }\textbf {\bibinfo {volume} {101}},\
  \bibinfo {pages} {124055} (\bibinfo {year} {2020})},\ \Eprint
  {https://arxiv.org/abs/2004.00009} {arXiv:2004.00009}\BibitemShut {NoStop}%
\bibitem [{\citenamefont {Silva}\ \emph {et~al.}(2021)\citenamefont {Silva},
  \citenamefont {Witek}, \citenamefont {Elley},\  and\ \citenamefont
  {Yunes}}]{Silva:2020omi}%
  \BibitemOpen
  \bibfield  {author} {\bibinfo {author} {\bibfnamefont {H.~O.}\ \bibnamefont
  {Silva}}, \bibinfo {author} {\bibfnamefont {H.}~\bibnamefont {Witek}},
  \bibinfo {author} {\bibfnamefont {M.}~\bibnamefont {Elley}},  and\ \bibinfo
  {author} {\bibfnamefont {N.}~\bibnamefont {Yunes}},\ }\bibfield  {title}
  {\bibinfo {title} {{Dynamical Descalarization in Binary Black Hole
  Mergers}},\ }\href {https://doi.org/10.1103/PhysRevLett.127.031101}
  {\bibfield  {journal} {\bibinfo  {journal} {Phys. Rev. Lett.}\ }\textbf
  {\bibinfo {volume} {127}},\ \bibinfo {pages} {031101} (\bibinfo {year}
  {2021})},\ \Eprint {https://arxiv.org/abs/2012.10436}
  {arXiv:2012.10436}\BibitemShut {NoStop}%
\bibitem [{\citenamefont {East}\  and\ \citenamefont
  {Ripley}(2021{\natexlab{a}})}]{East:2020hgw}%
  \BibitemOpen
  \bibfield  {author} {\bibinfo {author} {\bibfnamefont {W.~E.}\ \bibnamefont
  {East}} and\ \bibinfo {author} {\bibfnamefont {J.~L.}\ \bibnamefont
  {Ripley}},\ }\bibfield  {title} {\bibinfo {title} {{Evolution of
  Einstein-scalar-Gauss-Bonnet gravity using a modified harmonic
  formulation}},\ }\href {https://doi.org/10.1103/PhysRevD.103.044040}
  {\bibfield  {journal} {\bibinfo  {journal} {Phys. Rev. D}\ }\textbf {\bibinfo
  {volume} {103}},\ \bibinfo {pages} {044040} (\bibinfo {year}
  {2021}{\natexlab{a}})},\ \Eprint {https://arxiv.org/abs/2011.03547}
  {arXiv:2011.03547}\BibitemShut {NoStop}%
\bibitem [{\citenamefont {East}\  and\ \citenamefont
  {Ripley}(2021{\natexlab{b}})}]{East:2021bqk}%
  \BibitemOpen
  \bibfield  {author} {\bibinfo {author} {\bibfnamefont {W.~E.}\ \bibnamefont
  {East}} and\ \bibinfo {author} {\bibfnamefont {J.~L.}\ \bibnamefont
  {Ripley}},\ }\bibfield  {title} {\bibinfo {title} {{Dynamics of Spontaneous
  Black Hole Scalarization and Mergers in Einstein-Scalar-Gauss-Bonnet
  Gravity}},\ }\href {https://doi.org/10.1103/PhysRevLett.127.101102}
  {\bibfield  {journal} {\bibinfo  {journal} {Phys. Rev. Lett.}\ }\textbf
  {\bibinfo {volume} {127}},\ \bibinfo {pages} {101102} (\bibinfo {year}
  {2021}{\natexlab{b}})},\ \Eprint {https://arxiv.org/abs/2105.08571}
  {arXiv:2105.08571}\BibitemShut {NoStop}%
\bibitem [{\citenamefont {Doneva}\ \emph {et~al.}(2022)\citenamefont {Doneva},
  \citenamefont {Va\~n\'o Vi\~nuales},\  and\ \citenamefont
  {Yazadjiev}}]{Doneva:2022byd}%
  \BibitemOpen
  \bibfield  {author} {\bibinfo {author} {\bibfnamefont {D.~D.}\ \bibnamefont
  {Doneva}}, \bibinfo {author} {\bibfnamefont {A.}~\bibnamefont {Va\~n\'o
  Vi\~nuales}},  and\ \bibinfo {author} {\bibfnamefont {S.~S.}\ \bibnamefont
  {Yazadjiev}},\ }\bibfield  {title} {\bibinfo {title} {{Dynamical
  descalarization with a jump during black hole merger}},\ }\Eprint
  {https://arxiv.org/abs/2204.05333} {arXiv:2204.05333}\BibitemShut {NoStop}%
\bibitem [{\citenamefont {Okounkova}\ \emph {et~al.}(2017)\citenamefont
  {Okounkova}, \citenamefont {Stein}, \citenamefont {Scheel},\  and\
  \citenamefont {Hemberger}}]{Okounkova:2017yby}%
  \BibitemOpen
  \bibfield  {author} {\bibinfo {author} {\bibfnamefont {M.}~\bibnamefont
  {Okounkova}}, \bibinfo {author} {\bibfnamefont {L.~C.}\ \bibnamefont
  {Stein}}, \bibinfo {author} {\bibfnamefont {M.~A.}\ \bibnamefont {Scheel}},
  and\ \bibinfo {author} {\bibfnamefont {D.~A.}\ \bibnamefont {Hemberger}},\
  }\bibfield  {title} {\bibinfo {title} {{Numerical binary black hole mergers
  in dynamical Chern-Simons gravity: Scalar field}},\ }\href
  {https://doi.org/10.1103/PhysRevD.96.044020} {\bibfield  {journal} {\bibinfo
  {journal} {Phys. Rev. D}\ }\textbf {\bibinfo {volume} {96}},\ \bibinfo
  {pages} {044020} (\bibinfo {year} {2017})},\ \Eprint
  {https://arxiv.org/abs/1705.07924} {arXiv:1705.07924}\BibitemShut {NoStop}%
\bibitem [{\citenamefont {Okounkova}\ \emph {et~al.}(2019)\citenamefont
  {Okounkova}, \citenamefont {Stein}, \citenamefont {Scheel},\  and\
  \citenamefont {Teukolsky}}]{Okounkova:2019dfo}%
  \BibitemOpen
  \bibfield  {author} {\bibinfo {author} {\bibfnamefont {M.}~\bibnamefont
  {Okounkova}}, \bibinfo {author} {\bibfnamefont {L.~C.}\ \bibnamefont
  {Stein}}, \bibinfo {author} {\bibfnamefont {M.~A.}\ \bibnamefont {Scheel}},
  and\ \bibinfo {author} {\bibfnamefont {S.~A.}\ \bibnamefont {Teukolsky}},\
  }\bibfield  {title} {\bibinfo {title} {{Numerical binary black hole
  collisions in dynamical Chern-Simons gravity}},\ }\href
  {https://doi.org/10.1103/PhysRevD.100.104026} {\bibfield  {journal} {\bibinfo
   {journal} {Phys. Rev. D}\ }\textbf {\bibinfo {volume} {100}},\ \bibinfo
  {pages} {104026} (\bibinfo {year} {2019})},\ \Eprint
  {https://arxiv.org/abs/1906.08789} {arXiv:1906.08789}\BibitemShut {NoStop}%
\bibitem [{\citenamefont {Okounkova}\ \emph {et~al.}(2020)\citenamefont
  {Okounkova}, \citenamefont {Stein}, \citenamefont {Moxon}, \citenamefont
  {Scheel},\  and\ \citenamefont {Teukolsky}}]{Okounkova:2019zjf}%
  \BibitemOpen
  \bibfield  {author} {\bibinfo {author} {\bibfnamefont {M.}~\bibnamefont
  {Okounkova}}, \bibinfo {author} {\bibfnamefont {L.~C.}\ \bibnamefont
  {Stein}}, \bibinfo {author} {\bibfnamefont {J.}~\bibnamefont {Moxon}},
  \bibinfo {author} {\bibfnamefont {M.~A.}\ \bibnamefont {Scheel}},  and\
  \bibinfo {author} {\bibfnamefont {S.~A.}\ \bibnamefont {Teukolsky}},\
  }\bibfield  {title} {\bibinfo {title} {{Numerical relativity simulation of
  GW150914 beyond general relativity}},\ }\href
  {https://doi.org/10.1103/PhysRevD.101.104016} {\bibfield  {journal} {\bibinfo
   {journal} {Phys. Rev. D}\ }\textbf {\bibinfo {volume} {101}},\ \bibinfo
  {pages} {104016} (\bibinfo {year} {2020})},\ \Eprint
  {https://arxiv.org/abs/1911.02588} {arXiv:1911.02588}\BibitemShut {NoStop}%
\bibitem [{\citenamefont {Okounkova}(2020)}]{Okounkova:2020rqw}%
  \BibitemOpen
  \bibfield  {author} {\bibinfo {author} {\bibfnamefont {M.}~\bibnamefont
  {Okounkova}},\ }\bibfield  {title} {\bibinfo {title} {{Numerical relativity
  simulation of GW150914 in Einstein dilaton Gauss-Bonnet gravity}},\ }\href
  {https://doi.org/10.1103/PhysRevD.102.084046} {\bibfield  {journal} {\bibinfo
   {journal} {Phys. Rev. D}\ }\textbf {\bibinfo {volume} {102}},\ \bibinfo
  {pages} {084046} (\bibinfo {year} {2020})},\ \Eprint
  {https://arxiv.org/abs/2001.03571} {arXiv:2001.03571}\BibitemShut {NoStop}%
\bibitem [{\citenamefont {Bonilla}\ \emph {et~al.}(2022)\citenamefont
  {Bonilla}, \citenamefont {Kumar},\  and\ \citenamefont
  {Teukolsky}}]{Bonilla:2022dyt}%
  \BibitemOpen
  \bibfield  {author} {\bibinfo {author} {\bibfnamefont {G.~S.}\ \bibnamefont
  {Bonilla}}, \bibinfo {author} {\bibfnamefont {P.}~\bibnamefont {Kumar}},
  and\ \bibinfo {author} {\bibfnamefont {S.~A.}\ \bibnamefont {Teukolsky}},\
  }\bibfield  {title} {\bibinfo {title} {Modeling compact binary merger
  waveforms beyond general relativity},\ }\Eprint
  {https://arxiv.org/abs/2203.14026} {arXiv:2203.14026}\BibitemShut {NoStop}%
\bibitem [{\citenamefont {Pratten}\ \emph {et~al.}(2020)\citenamefont
  {Pratten}, \citenamefont {Husa}, \citenamefont {Garcia-Quiros}, \citenamefont
  {Colleoni}, \citenamefont {Ramos-Buades}, \citenamefont {Estelles},\  and\
  \citenamefont {Jaume}}]{Pratten:2020fqn}%
  \BibitemOpen
  \bibfield  {author} {\bibinfo {author} {\bibfnamefont {G.}~\bibnamefont
  {Pratten}}, \bibinfo {author} {\bibfnamefont {S.}~\bibnamefont {Husa}},
  \bibinfo {author} {\bibfnamefont {C.}~\bibnamefont {Garcia-Quiros}}, \bibinfo
  {author} {\bibfnamefont {M.}~\bibnamefont {Colleoni}}, \bibinfo {author}
  {\bibfnamefont {A.}~\bibnamefont {Ramos-Buades}}, \bibinfo {author}
  {\bibfnamefont {H.}~\bibnamefont {Estelles}},  and\ \bibinfo {author}
  {\bibfnamefont {R.}~\bibnamefont {Jaume}},\ }\bibfield  {title} {\bibinfo
  {title} {{Setting the cornerstone for a family of models for gravitational
  waves from compact binaries: The dominant harmonic for nonprecessing
  quasicircular black holes}},\ }\href
  {https://doi.org/10.1103/PhysRevD.102.064001} {\bibfield  {journal} {\bibinfo
   {journal} {Phys. Rev. D}\ }\textbf {\bibinfo {volume} {102}},\ \bibinfo
  {pages} {064001} (\bibinfo {year} {2020})},\ \Eprint
  {https://arxiv.org/abs/2001.11412} {arXiv:2001.11412}\BibitemShut {NoStop}%
\bibitem [{\citenamefont {Garc\'{\i}a-Quir\'os}\ \emph
  {et~al.}(2020)\citenamefont {Garc\'{\i}a-Quir\'os}, \citenamefont {Colleoni},
  \citenamefont {Husa}, \citenamefont {Estell\'es}, \citenamefont {Pratten},
  \citenamefont {Ramos-Buades}, \citenamefont {Mateu-Lucena},\  and\
  \citenamefont {Jaume}}]{Garcia-Quiros:2020qpx}%
  \BibitemOpen
  \bibfield  {author} {\bibinfo {author} {\bibfnamefont {C.}~\bibnamefont
  {Garc\'{\i}a-Quir\'os}}, \bibinfo {author} {\bibfnamefont {M.}~\bibnamefont
  {Colleoni}}, \bibinfo {author} {\bibfnamefont {S.}~\bibnamefont {Husa}},
  \bibinfo {author} {\bibfnamefont {H.}~\bibnamefont {Estell\'es}}, \bibinfo
  {author} {\bibfnamefont {G.}~\bibnamefont {Pratten}}, \bibinfo {author}
  {\bibfnamefont {A.}~\bibnamefont {Ramos-Buades}}, \bibinfo {author}
  {\bibfnamefont {M.}~\bibnamefont {Mateu-Lucena}},  and\ \bibinfo {author}
  {\bibfnamefont {R.}~\bibnamefont {Jaume}},\ }\bibfield  {title} {\bibinfo
  {title} {Multimode frequency-domain model for the gravitational wave signal
  from nonprecessing black-hole binaries},\ }\href
  {https://doi.org/10.1103/PhysRevD.102.064002} {\bibfield  {journal} {\bibinfo
   {journal} {Phys. Rev. D}\ }\textbf {\bibinfo {volume} {102}},\ \bibinfo
  {pages} {064002} (\bibinfo {year} {2020})},\ \Eprint
  {https://arxiv.org/abs/2001.10914} {arXiv:2001.10914}\BibitemShut {NoStop}%
\bibitem [{\citenamefont {Pratten}\ \emph {et~al.}(2021)\citenamefont
  {Pratten}, \citenamefont {Garc\'{\i}a-Quir\'os}, \citenamefont {Colleoni},
  \citenamefont {Ramos-Buades}, \citenamefont {Estell\'es}, \citenamefont
  {Mateu-Lucena}, \citenamefont {Jaume}, \citenamefont {Haney}, \citenamefont
  {Keitel}, \citenamefont {Thompson},\  and\ \citenamefont
  {Husa}}]{Pratten:2020ceb}%
  \BibitemOpen
  \bibfield  {author} {\bibinfo {author} {\bibfnamefont {G.}~\bibnamefont
  {Pratten}}, \bibinfo {author} {\bibfnamefont {C.}~\bibnamefont
  {Garc\'{\i}a-Quir\'os}}, \bibinfo {author} {\bibfnamefont {M.}~\bibnamefont
  {Colleoni}}, \bibinfo {author} {\bibfnamefont {A.}~\bibnamefont
  {Ramos-Buades}}, \bibinfo {author} {\bibfnamefont {H.}~\bibnamefont
  {Estell\'es}}, \bibinfo {author} {\bibfnamefont {M.}~\bibnamefont
  {Mateu-Lucena}}, \bibinfo {author} {\bibfnamefont {R.}~\bibnamefont {Jaume}},
  \bibinfo {author} {\bibfnamefont {M.}~\bibnamefont {Haney}}, \bibinfo
  {author} {\bibfnamefont {D.}~\bibnamefont {Keitel}}, \bibinfo {author}
  {\bibfnamefont {J.~E.}\ \bibnamefont {Thompson}},  and\ \bibinfo {author}
  {\bibfnamefont {S.}~\bibnamefont {Husa}},\ }\bibfield  {title} {\bibinfo
  {title} {Computationally efficient models for the dominant and subdominant
  harmonic modes of precessing binary black holes},\ }\href
  {https://doi.org/10.1103/PhysRevD.103.104056} {\bibfield  {journal} {\bibinfo
   {journal} {Phys. Rev. D}\ }\textbf {\bibinfo {volume} {103}},\ \bibinfo
  {pages} {104056} (\bibinfo {year} {2021})},\ \Eprint
  {https://arxiv.org/abs/2004.06503} {arXiv:2004.06503}\BibitemShut {NoStop}%
\bibitem [{\citenamefont {Ashton}\ \emph {et~al.}(2019)\citenamefont {Ashton},
  \citenamefont {H{\"u}bner}, \citenamefont {Lasky}, \citenamefont {Talbot},
  \citenamefont {Ackley}, \citenamefont {Biscoveanu}, \citenamefont {Chu},
  \citenamefont {Divakarla}, \citenamefont {Easter}, \citenamefont {Goncharov},
  \citenamefont {Vivanco}, \citenamefont {Harms}, \citenamefont {Lower},
  \citenamefont {Meadors}, \citenamefont {Melchor} \emph
  {et~al.}}]{Ashton:2018jfp}%
  \BibitemOpen
  \bibfield  {author} {\bibinfo {author} {\bibfnamefont {G.}~\bibnamefont
  {Ashton}}, \bibinfo {author} {\bibfnamefont {M.}~\bibnamefont {H{\"u}bner}},
  \bibinfo {author} {\bibfnamefont {P.~D.}\ \bibnamefont {Lasky}}, \bibinfo
  {author} {\bibfnamefont {C.}~\bibnamefont {Talbot}}, \bibinfo {author}
  {\bibfnamefont {K.}~\bibnamefont {Ackley}}, \bibinfo {author} {\bibfnamefont
  {S.}~\bibnamefont {Biscoveanu}}, \bibinfo {author} {\bibfnamefont
  {Q.}~\bibnamefont {Chu}}, \bibinfo {author} {\bibfnamefont {A.}~\bibnamefont
  {Divakarla}}, \bibinfo {author} {\bibfnamefont {P.~J.}\ \bibnamefont
  {Easter}}, \bibinfo {author} {\bibfnamefont {B.}~\bibnamefont {Goncharov}},
  \bibinfo {author} {\bibfnamefont {F.~H.}\ \bibnamefont {Vivanco}}, \bibinfo
  {author} {\bibfnamefont {J.}~\bibnamefont {Harms}}, \bibinfo {author}
  {\bibfnamefont {M.~E.}\ \bibnamefont {Lower}}, \bibinfo {author}
  {\bibfnamefont {G.~D.}\ \bibnamefont {Meadors}}, \bibinfo {author}
  {\bibfnamefont {D.}~\bibnamefont {Melchor}},  \emph {et~al.},\ }\bibfield
  {title} {\bibinfo {title} {{BILBY: A user-friendly Bayesian inference library
  for gravitational-wave astronomy}},\ }\href
  {https://doi.org/10.3847/1538-4365/ab06fc} {\bibfield  {journal} {\bibinfo
  {journal} {Astrophys. J. Suppl. Ser.}\ }\textbf {\bibinfo {volume} {241}},\
  \bibinfo {pages} {27} (\bibinfo {year} {2019})},\ \Eprint
  {https://arxiv.org/abs/1811.02042} {arXiv:1811.02042}\BibitemShut {NoStop}%
\bibitem [{\citenamefont {Romero-Shaw}\ \emph {et~al.}(2020)\citenamefont
  {Romero-Shaw}, \citenamefont {Talbot}, \citenamefont {Biscoveanu},
  \citenamefont {D'Emilio}, \citenamefont {Ashton}, \citenamefont {Berry},
  \citenamefont {Coughlin}, \citenamefont {Galaudage}, \citenamefont {Hoy},
  \citenamefont {H{\"u}bner}, \citenamefont {Phukon}, \citenamefont {Pitkin},
  \citenamefont {Rizzo}, \citenamefont {Sarin}, \citenamefont {Smith} \emph
  {et~al.}}]{Romero-Shaw:2020owr}%
  \BibitemOpen
  \bibfield  {author} {\bibinfo {author} {\bibfnamefont {I.~M.}\ \bibnamefont
  {Romero-Shaw}}, \bibinfo {author} {\bibfnamefont {C.}~\bibnamefont {Talbot}},
  \bibinfo {author} {\bibfnamefont {S.}~\bibnamefont {Biscoveanu}}, \bibinfo
  {author} {\bibfnamefont {V.}~\bibnamefont {D'Emilio}}, \bibinfo {author}
  {\bibfnamefont {G.}~\bibnamefont {Ashton}}, \bibinfo {author} {\bibfnamefont
  {C.~P.~L.}\ \bibnamefont {Berry}}, \bibinfo {author} {\bibfnamefont
  {S.}~\bibnamefont {Coughlin}}, \bibinfo {author} {\bibfnamefont
  {S.}~\bibnamefont {Galaudage}}, \bibinfo {author} {\bibfnamefont
  {C.}~\bibnamefont {Hoy}}, \bibinfo {author} {\bibfnamefont {M.}~\bibnamefont
  {H{\"u}bner}}, \bibinfo {author} {\bibfnamefont {K.~S.}\ \bibnamefont
  {Phukon}}, \bibinfo {author} {\bibfnamefont {M.}~\bibnamefont {Pitkin}},
  \bibinfo {author} {\bibfnamefont {M.}~\bibnamefont {Rizzo}}, \bibinfo
  {author} {\bibfnamefont {N.}~\bibnamefont {Sarin}}, \bibinfo {author}
  {\bibfnamefont {R.}~\bibnamefont {Smith}},  \emph {et~al.},\ }\bibfield
  {title} {\bibinfo {title} {{Bayesian inference for compact binary
  coalescences with BILBY: {V}alidation and application to the first
  LIGO\textendash{}Virgo gravitational-wave transient catalogue}},\ }\href
  {https://doi.org/10.1093/mnras/staa2850} {\bibfield  {journal} {\bibinfo
  {journal} {Mon. Not. R. Astron. Soc.}\ }\textbf {\bibinfo {volume} {499}},\
  \bibinfo {pages} {3295} (\bibinfo {year} {2020})},\ \Eprint
  {https://arxiv.org/abs/2006.00714} {arXiv:2006.00714}\BibitemShut {NoStop}%
\bibitem [{\citenamefont {Abbott}\ \emph
  {et~al.}(2021{\natexlab{c}})\citenamefont {Abbott} \emph
  {et~al.}}]{LIGOScientific:2019lzm}%
  \BibitemOpen
  \bibfield  {author} {\bibinfo {author} {\bibfnamefont {R.}~\bibnamefont
  {Abbott}} \emph {et~al.} (\bibinfo {collaboration} {LIGO Scientific and Virgo
  Collaborations}),\ }\bibfield  {title} {\bibinfo {title} {{Open data from the
  first and second observing runs of Advanced LIGO and Advanced Virgo}},\
  }\href {https://doi.org/10.1016/j.softx.2021.100658} {\bibfield  {journal}
  {\bibinfo  {journal} {SoftwareX}\ }\textbf {\bibinfo {volume} {13}},\
  \bibinfo {pages} {100658} (\bibinfo {year} {2021}{\natexlab{c}})},\ \Eprint
  {https://arxiv.org/abs/1912.11716} {arXiv:1912.11716}\BibitemShut {NoStop}%
\bibitem [{\citenamefont {Chatziioannou}\ \emph {et~al.}(2019)\citenamefont
  {Chatziioannou}, \citenamefont {Haster}, \citenamefont {Littenberg},
  \citenamefont {Farr}, \citenamefont {Ghonge}, \citenamefont {Millhouse},
  \citenamefont {Clark},\  and\ \citenamefont
  {Cornish}}]{Chatziioannou:2019zvs}%
  \BibitemOpen
  \bibfield  {author} {\bibinfo {author} {\bibfnamefont {K.}~\bibnamefont
  {Chatziioannou}}, \bibinfo {author} {\bibfnamefont {C.-J.}\ \bibnamefont
  {Haster}}, \bibinfo {author} {\bibfnamefont {T.~B.}\ \bibnamefont
  {Littenberg}}, \bibinfo {author} {\bibfnamefont {W.~M.}\ \bibnamefont
  {Farr}}, \bibinfo {author} {\bibfnamefont {S.}~\bibnamefont {Ghonge}},
  \bibinfo {author} {\bibfnamefont {M.}~\bibnamefont {Millhouse}}, \bibinfo
  {author} {\bibfnamefont {J.~A.}\ \bibnamefont {Clark}},  and\ \bibinfo
  {author} {\bibfnamefont {N.}~\bibnamefont {Cornish}},\ }\bibfield  {title}
  {\bibinfo {title} {{Noise spectral estimation methods and their impact on
  gravitational wave measurement of compact binary mergers}},\ }\href
  {https://doi.org/10.1103/PhysRevD.100.104004} {\bibfield  {journal} {\bibinfo
   {journal} {Phys. Rev. D}\ }\textbf {\bibinfo {volume} {100}},\ \bibinfo
  {pages} {104004} (\bibinfo {year} {2019})},\ \Eprint
  {https://arxiv.org/abs/1907.06540} {arXiv:1907.06540}\BibitemShut {NoStop}%
\bibitem [{\citenamefont {Abbott}\ \emph
  {et~al.}(2019{\natexlab{c}})\citenamefont {Abbott} \emph {et~al.}}]{gwtc1}%
  \BibitemOpen
  \bibfield  {author} {\bibinfo {author} {\bibfnamefont {B.~P.}\ \bibnamefont
  {Abbott}} \emph {et~al.} (\bibinfo {collaboration} {LIGO Scientific and Virgo
  Collaborations}),\ }\bibfield  {title} {\bibinfo {title} {{GWTC-1: A
  Gravitational-Wave Transient Catalog of Compact Binary Mergers Observed by
  LIGO and Virgo during the First and Second Observing Runs}},\ }\href
  {https://doi.org/10.1103/PhysRevX.9.031040} {\bibfield  {journal} {\bibinfo
  {journal} {Phys. Rev. X}\ }\textbf {\bibinfo {volume} {9}},\ \bibinfo {pages}
  {031040} (\bibinfo {year} {2019}{\natexlab{c}})},\ \Eprint
  {https://arxiv.org/abs/1811.12907} {arXiv:1811.12907}\BibitemShut {NoStop}%
\bibitem [{\citenamefont {Ade}\ \emph {et~al.}(2016)\citenamefont {Ade} \emph
  {et~al.}}]{Planck:2015fie}%
  \BibitemOpen
  \bibfield  {author} {\bibinfo {author} {\bibfnamefont {P.~A.~R.}\
  \bibnamefont {Ade}} \emph {et~al.} (\bibinfo {collaboration} {Planck
  Collaboration}),\ }\bibfield  {title} {\bibinfo {title} {{Planck 2015
  results. XIII. Cosmological parameters}},\ }\href
  {https://doi.org/10.1051/0004-6361/201525830} {\bibfield  {journal} {\bibinfo
   {journal} {Astron. Astrophys.}\ }\textbf {\bibinfo {volume} {594}},\
  \bibinfo {pages} {A13} (\bibinfo {year} {2016})},\ \Eprint
  {https://arxiv.org/abs/1502.01589} {arXiv:1502.01589}\BibitemShut {NoStop}%
\bibitem [{\citenamefont {{Speagle}}(2020)}]{dynesty}%
  \BibitemOpen
  \bibfield  {author} {\bibinfo {author} {\bibfnamefont {J.~S.}\ \bibnamefont
  {{Speagle}}},\ }\bibfield  {title} {\bibinfo {title} {{DYNESTY: A dynamic
  nested sampling package for estimating Bayesian posteriors and evidences}},\
  }\href {https://doi.org/10.1093/mnras/staa278} {\bibfield  {journal}
  {\bibinfo  {journal} {Mon. Not. R. Astron. Soc.}\ }\textbf {\bibinfo {volume}
  {493}},\ \bibinfo {pages} {3132} (\bibinfo {year} {2020})},\ \Eprint
  {https://arxiv.org/abs/1904.02180} {arXiv:1904.02180}\BibitemShut {NoStop}%
\bibitem [{\citenamefont {Skilling}(2004)}]{Skilling:2004nsp}%
  \BibitemOpen
  \bibfield  {author} {\bibinfo {author} {\bibfnamefont {J.}~\bibnamefont
  {Skilling}},\ }\bibfield  {title} {\bibinfo {title} {Nested sampling},\
  }\href {https://doi.org/10.1063/1.1835238} {\bibfield  {journal} {\bibinfo
  {journal} {AIP Conf. Proc.}\ }\textbf {\bibinfo {volume} {735}},\ \bibinfo
  {pages} {395} (\bibinfo {year} {2004})}\BibitemShut {NoStop}%
\bibitem [{\citenamefont {Skilling}(2006)}]{Skilling:2006nsp}%
  \BibitemOpen
  \bibfield  {author} {\bibinfo {author} {\bibfnamefont {J.}~\bibnamefont
  {Skilling}},\ }\bibfield  {title} {\bibinfo {title} {{Nested sampling for
  general Bayesian computation}},\ }\href {https://doi.org/10.1214/06-BA127}
  {\bibfield  {journal} {\bibinfo  {journal} {Bayesian Anal.}\ }\textbf
  {\bibinfo {volume} {1}},\ \bibinfo {pages} {833 } (\bibinfo {year}
  {2006})}\BibitemShut {NoStop}%
\bibitem [{\citenamefont {{Higson}}\ \emph {et~al.}(2019)\citenamefont
  {{Higson}}, \citenamefont {{Handley}}, \citenamefont {{Hobson}},\  and\
  \citenamefont {{Lasenby}}}]{Higson:2019dns}%
  \BibitemOpen
  \bibfield  {author} {\bibinfo {author} {\bibfnamefont {E.}~\bibnamefont
  {{Higson}}}, \bibinfo {author} {\bibfnamefont {W.}~\bibnamefont {{Handley}}},
  \bibinfo {author} {\bibfnamefont {M.}~\bibnamefont {{Hobson}}},  and\
  \bibinfo {author} {\bibfnamefont {A.}~\bibnamefont {{Lasenby}}},\ }\bibfield
  {title} {\bibinfo {title} {{Dynamic nested sampling: An improved algorithm
  for parameter estimation and evidence calculation}},\ }\href
  {https://doi.org/10.1007/s11222-018-9844-0} {\bibfield  {journal} {\bibinfo
  {journal} {Stat. Comput.}\ }\textbf {\bibinfo {volume} {29}},\ \bibinfo
  {pages} {891} (\bibinfo {year} {2019})},\ \Eprint
  {https://arxiv.org/abs/1704.03459} {arXiv:1704.03459}\BibitemShut {NoStop}%
\bibitem [{\citenamefont {Abbott}\ \emph
  {et~al.}(2021{\natexlab{d}})\citenamefont {Abbott} \emph {et~al.}}]{gwtc2}%
  \BibitemOpen
  \bibfield  {author} {\bibinfo {author} {\bibfnamefont {R.}~\bibnamefont
  {Abbott}} \emph {et~al.} (\bibinfo {collaboration} {LIGO Scientific and Virgo
  Collaborations}),\ }\bibfield  {title} {\bibinfo {title} {{GWTC-2: Compact
  Binary Coalescences Observed by LIGO and Virgo During the First Half of the
  Third Observing Run}},\ }\href {https://doi.org/10.1103/PhysRevX.11.021053}
  {\bibfield  {journal} {\bibinfo  {journal} {Phys. Rev. X}\ }\textbf {\bibinfo
  {volume} {11}},\ \bibinfo {pages} {021053} (\bibinfo {year}
  {2021}{\natexlab{d}})},\ \Eprint {https://arxiv.org/abs/2010.14527}
  {arXiv:2010.14527}\BibitemShut {NoStop}%
\bibitem [{\citenamefont {Abbott}\ \emph
  {et~al.}(2021{\natexlab{e}})\citenamefont {Abbott} \emph {et~al.}}]{gwtc3}%
  \BibitemOpen
  \bibfield  {author} {\bibinfo {author} {\bibfnamefont {R.}~\bibnamefont
  {Abbott}} \emph {et~al.} (\bibinfo {collaboration} {LIGO Scientific, Virgo,
  and KAGRA Collaborations}),\ }\bibfield  {title} {\bibinfo {title} {{GWTC-3:
  C}ompact binary coalescences observed by {LIGO} and {Virgo} during the second
  part of the third observing run},\ }\Eprint
  {https://arxiv.org/abs/2111.03606} {arXiv:2111.03606}\BibitemShut {NoStop}%
\bibitem [{\citenamefont {Doneva}\  and\ \citenamefont
  {Yazadjiev}(2018{\natexlab{b}})}]{Doneva:2017duq}%
  \BibitemOpen
  \bibfield  {author} {\bibinfo {author} {\bibfnamefont {D.~D.}\ \bibnamefont
  {Doneva}} and\ \bibinfo {author} {\bibfnamefont {S.~S.}\ \bibnamefont
  {Yazadjiev}},\ }\bibfield  {title} {\bibinfo {title} {{Neutron star solutions
  with curvature induced scalarization in the extended Gauss-Bonnet
  scalar-tensor theories}},\ }\href
  {https://doi.org/10.1088/1475-7516/2018/04/011} {\bibfield  {journal}
  {\bibinfo  {journal} {J. Cosmol. Astropart. Phys.}\ }04 (2018)\ 011},\
  \Eprint {https://arxiv.org/abs/1712.03715} {arXiv:1712.03715}\BibitemShut
  {NoStop}%
\bibitem [{\citenamefont {Ventagli}\ \emph {et~al.}(2021)\citenamefont
  {Ventagli}, \citenamefont {Antoniou}, \citenamefont {Leh\'ebel},\  and\
  \citenamefont {Sotiriou}}]{Ventagli:2021ubn}%
  \BibitemOpen
  \bibfield  {author} {\bibinfo {author} {\bibfnamefont {G.}~\bibnamefont
  {Ventagli}}, \bibinfo {author} {\bibfnamefont {G.}~\bibnamefont {Antoniou}},
  \bibinfo {author} {\bibfnamefont {A.}~\bibnamefont {Leh\'ebel}},  and\
  \bibinfo {author} {\bibfnamefont {T.~P.}\ \bibnamefont {Sotiriou}},\
  }\bibfield  {title} {\bibinfo {title} {{Neutron star scalarization with
  Gauss-Bonnet and Ricci scalar couplings}},\ }\href
  {https://doi.org/10.1103/PhysRevD.104.124078} {\bibfield  {journal} {\bibinfo
   {journal} {Phys. Rev. D}\ }\textbf {\bibinfo {volume} {104}},\ \bibinfo
  {pages} {124078} (\bibinfo {year} {2021})},\ \Eprint
  {https://arxiv.org/abs/2111.03644} {arXiv:2111.03644}\BibitemShut {NoStop}%
\bibitem [{\citenamefont {Danchev}\ \emph {et~al.}(2021)\citenamefont
  {Danchev}, \citenamefont {Doneva},\  and\ \citenamefont
  {Yazadjiev}}]{Danchev:2021tew}%
  \BibitemOpen
  \bibfield  {author} {\bibinfo {author} {\bibfnamefont {V.~I.}\ \bibnamefont
  {Danchev}}, \bibinfo {author} {\bibfnamefont {D.~D.}\ \bibnamefont {Doneva}},
   and\ \bibinfo {author} {\bibfnamefont {S.~S.}\ \bibnamefont {Yazadjiev}},\
  }\bibfield  {title} {\bibinfo {title} {{Constraining scalarization in
  scalar-Gauss-Bonnet gravity through binary pulsars}},\ }\Eprint
  {https://arxiv.org/abs/2112.03869} {arXiv:2112.03869}\BibitemShut {NoStop}%
\bibitem [{\citenamefont {Dima}\ \emph {et~al.}(2020)\citenamefont {Dima},
  \citenamefont {Barausse}, \citenamefont {Franchini},\  and\ \citenamefont
  {Sotiriou}}]{Dima:2020yac}%
  \BibitemOpen
  \bibfield  {author} {\bibinfo {author} {\bibfnamefont {A.}~\bibnamefont
  {Dima}}, \bibinfo {author} {\bibfnamefont {E.}~\bibnamefont {Barausse}},
  \bibinfo {author} {\bibfnamefont {N.}~\bibnamefont {Franchini}},  and\
  \bibinfo {author} {\bibfnamefont {T.~P.}\ \bibnamefont {Sotiriou}},\
  }\bibfield  {title} {\bibinfo {title} {{Spin-Induced Black Hole Spontaneous
  Scalarization}},\ }\href {https://doi.org/10.1103/PhysRevLett.125.231101}
  {\bibfield  {journal} {\bibinfo  {journal} {Phys. Rev. Lett.}\ }\textbf
  {\bibinfo {volume} {125}},\ \bibinfo {pages} {231101} (\bibinfo {year}
  {2020})},\ \Eprint {https://arxiv.org/abs/2006.03095}
  {arXiv:2006.03095}\BibitemShut {NoStop}%
\bibitem [{\citenamefont {Herdeiro}\ \emph {et~al.}(2021)\citenamefont
  {Herdeiro}, \citenamefont {Radu}, \citenamefont {Silva}, \citenamefont
  {Sotiriou},\  and\ \citenamefont {Yunes}}]{Herdeiro:2020wei}%
  \BibitemOpen
  \bibfield  {author} {\bibinfo {author} {\bibfnamefont {C.~A.~R.}\
  \bibnamefont {Herdeiro}}, \bibinfo {author} {\bibfnamefont {E.}~\bibnamefont
  {Radu}}, \bibinfo {author} {\bibfnamefont {H.~O.}\ \bibnamefont {Silva}},
  \bibinfo {author} {\bibfnamefont {T.~P.}\ \bibnamefont {Sotiriou}},  and\
  \bibinfo {author} {\bibfnamefont {N.}~\bibnamefont {Yunes}},\ }\bibfield
  {title} {\bibinfo {title} {{Spin-Induced Scalarized Black Holes}},\ }\href
  {https://doi.org/10.1103/PhysRevLett.126.011103} {\bibfield  {journal}
  {\bibinfo  {journal} {Phys. Rev. Lett.}\ }\textbf {\bibinfo {volume} {126}},\
  \bibinfo {pages} {011103} (\bibinfo {year} {2021})},\ \Eprint
  {https://arxiv.org/abs/2009.03904} {arXiv:2009.03904}\BibitemShut {NoStop}%
\bibitem [{\citenamefont {Berti}\ \emph {et~al.}(2021)\citenamefont {Berti},
  \citenamefont {Collodel}, \citenamefont {Kleihaus},\  and\ \citenamefont
  {Kunz}}]{Berti:2020kgk}%
  \BibitemOpen
  \bibfield  {author} {\bibinfo {author} {\bibfnamefont {E.}~\bibnamefont
  {Berti}}, \bibinfo {author} {\bibfnamefont {L.~G.}\ \bibnamefont {Collodel}},
  \bibinfo {author} {\bibfnamefont {B.}~\bibnamefont {Kleihaus}},  and\
  \bibinfo {author} {\bibfnamefont {J.}~\bibnamefont {Kunz}},\ }\bibfield
  {title} {\bibinfo {title} {{Spin-Induced Black Hole Scalarization in
  Einstein-Scalar-Gauss-Bonnet Theory}},\ }\href
  {https://doi.org/10.1103/PhysRevLett.126.011104} {\bibfield  {journal}
  {\bibinfo  {journal} {Phys. Rev. Lett.}\ }\textbf {\bibinfo {volume} {126}},\
  \bibinfo {pages} {011104} (\bibinfo {year} {2021})},\ \Eprint
  {https://arxiv.org/abs/2009.03905} {arXiv:2009.03905}\BibitemShut {NoStop}%
\bibitem [{\citenamefont {Doneva}\ \emph {et~al.}(2020)\citenamefont {Doneva},
  \citenamefont {Collodel}, \citenamefont {Kr\"uger},\  and\ \citenamefont
  {Yazadjiev}}]{Doneva:2020kfv}%
  \BibitemOpen
  \bibfield  {author} {\bibinfo {author} {\bibfnamefont {D.~D.}\ \bibnamefont
  {Doneva}}, \bibinfo {author} {\bibfnamefont {L.~G.}\ \bibnamefont
  {Collodel}}, \bibinfo {author} {\bibfnamefont {C.~J.}\ \bibnamefont
  {Kr\"uger}},  and\ \bibinfo {author} {\bibfnamefont {S.~S.}\ \bibnamefont
  {Yazadjiev}},\ }\bibfield  {title} {\bibinfo {title} {{Spin-induced
  scalarization of Kerr black holes with a massive scalar field}},\ }\href
  {https://doi.org/10.1140/epjc/s10052-020-08765-3} {\bibfield  {journal}
  {\bibinfo  {journal} {Eur. Phys. J. C}\ }\textbf {\bibinfo {volume} {80}},\
  \bibinfo {pages} {1205} (\bibinfo {year} {2020})},\ \Eprint
  {https://arxiv.org/abs/2009.03774} {arXiv:2009.03774}\BibitemShut {NoStop}%
\bibitem [{\citenamefont {Hod}(2022)}]{Hod:2022hfm}%
  \BibitemOpen
  \bibfield  {author} {\bibinfo {author} {\bibfnamefont {S.}~\bibnamefont
  {Hod}},\ }\bibfield  {title} {\bibinfo {title} {{Spin-induced black hole
  spontaneous scalarization: Analytic treatment in the large-coupling
  regime}},\ }\href {https://doi.org/10.1103/PhysRevD.105.024074} {\bibfield
  {journal} {\bibinfo  {journal} {Phys. Rev. D}\ }\textbf {\bibinfo {volume}
  {105}},\ \bibinfo {pages} {024074} (\bibinfo {year} {2022})}\BibitemShut
  {NoStop}%
\bibitem [{gwo()}]{gwosc}%
  \BibitemOpen
  \href@noop {} {}\bibinfo {note}
  {\url{https://www.gw-openscience.org}}\BibitemShut {NoStop}%
\bibitem [{\citenamefont {{Robitaille}}\ \emph {et~al.}(2013)\citenamefont
  {{Robitaille}}, \citenamefont {{Tollerud}}, \citenamefont {{Greenfield}},
  \citenamefont {{Droettboom}}, \citenamefont {{Bray}}, \citenamefont
  {{Aldcroft}}, \citenamefont {{Davis}}, \citenamefont {{Ginsburg}},
  \citenamefont {{Price-Whelan}}, \citenamefont {{Kerzendorf}}, \citenamefont
  {{Conley}}, \citenamefont {{Crighton}}, \citenamefont {{Barbary}},
  \citenamefont {{Muna}}, \citenamefont {{Ferguson}} \emph
  {et~al.}}]{astropy:2013}%
  \BibitemOpen
  \bibfield  {author} {\bibinfo {author} {\bibfnamefont {T.~P.}\ \bibnamefont
  {{Robitaille}}}, \bibinfo {author} {\bibfnamefont {E.~J.}\ \bibnamefont
  {{Tollerud}}}, \bibinfo {author} {\bibfnamefont {P.}~\bibnamefont
  {{Greenfield}}}, \bibinfo {author} {\bibfnamefont {M.}~\bibnamefont
  {{Droettboom}}}, \bibinfo {author} {\bibfnamefont {E.}~\bibnamefont
  {{Bray}}}, \bibinfo {author} {\bibfnamefont {T.}~\bibnamefont {{Aldcroft}}},
  \bibinfo {author} {\bibfnamefont {M.}~\bibnamefont {{Davis}}}, \bibinfo
  {author} {\bibfnamefont {A.}~\bibnamefont {{Ginsburg}}}, \bibinfo {author}
  {\bibfnamefont {A.~M.}\ \bibnamefont {{Price-Whelan}}}, \bibinfo {author}
  {\bibfnamefont {W.~E.}\ \bibnamefont {{Kerzendorf}}}, \bibinfo {author}
  {\bibfnamefont {A.}~\bibnamefont {{Conley}}}, \bibinfo {author}
  {\bibfnamefont {N.}~\bibnamefont {{Crighton}}}, \bibinfo {author}
  {\bibfnamefont {K.}~\bibnamefont {{Barbary}}}, \bibinfo {author}
  {\bibfnamefont {D.}~\bibnamefont {{Muna}}}, \bibinfo {author} {\bibfnamefont
  {H.}~\bibnamefont {{Ferguson}}},  \emph {et~al.} (\bibinfo {collaboration}
  {Astropy Collaboration}),\ }\bibfield  {title} {\bibinfo {title} {{Astropy: A
  community Python package for astronomy}},\ }\href
  {https://doi.org/10.1051/0004-6361/201322068} {\bibfield  {journal} {\bibinfo
   {journal} {Astron. \& Astrophys.}\ }\textbf {\bibinfo {volume} {558}},\
  \bibinfo {pages} {A33} (\bibinfo {year} {2013})},\ \Eprint
  {https://arxiv.org/abs/1307.6212} {arXiv:1307.6212}\BibitemShut {NoStop}%
\bibitem [{\citenamefont {{Price-Whelan}}\ \emph {et~al.}(2018)\citenamefont
  {{Price-Whelan}}, \citenamefont {{Sip{\H{o}}cz}}, \citenamefont
  {{G{\"u}nther}}, \citenamefont {{Lim}}, \citenamefont {{Crawford}},
  \citenamefont {{Conseil}}, \citenamefont {{Shupe}}, \citenamefont {{Craig}},
  \citenamefont {{Dencheva}}, \citenamefont {{Ginsburg}}, \citenamefont
  {{VanderPlas}}, \citenamefont {{Bradley}}, \citenamefont
  {{P{\'e}rez-Su{\'a}rez}}, \citenamefont {{de Val-Borro}}, \citenamefont
  {{Aldcroft}} \emph {et~al.}}]{astropy:2018}%
  \BibitemOpen
  \bibfield  {author} {\bibinfo {author} {\bibfnamefont {A.~M.}\ \bibnamefont
  {{Price-Whelan}}}, \bibinfo {author} {\bibfnamefont {B.~M.}\ \bibnamefont
  {{Sip{\H{o}}cz}}}, \bibinfo {author} {\bibfnamefont {H.~M.}\ \bibnamefont
  {{G{\"u}nther}}}, \bibinfo {author} {\bibfnamefont {P.~L.}\ \bibnamefont
  {{Lim}}}, \bibinfo {author} {\bibfnamefont {S.~M.}\ \bibnamefont
  {{Crawford}}}, \bibinfo {author} {\bibfnamefont {S.}~\bibnamefont
  {{Conseil}}}, \bibinfo {author} {\bibfnamefont {D.~L.}\ \bibnamefont
  {{Shupe}}}, \bibinfo {author} {\bibfnamefont {M.~W.}\ \bibnamefont
  {{Craig}}}, \bibinfo {author} {\bibfnamefont {N.}~\bibnamefont {{Dencheva}}},
  \bibinfo {author} {\bibfnamefont {A.}~\bibnamefont {{Ginsburg}}}, \bibinfo
  {author} {\bibfnamefont {J.~T.}\ \bibnamefont {{VanderPlas}}}, \bibinfo
  {author} {\bibfnamefont {L.~D.}\ \bibnamefont {{Bradley}}}, \bibinfo {author}
  {\bibfnamefont {D.}~\bibnamefont {{P{\'e}rez-Su{\'a}rez}}}, \bibinfo {author}
  {\bibfnamefont {M.}~\bibnamefont {{de Val-Borro}}}, \bibinfo {author}
  {\bibfnamefont {T.~L.}\ \bibnamefont {{Aldcroft}}},  \emph {et~al.} (\bibinfo
  {collaboration} {Astropy Collaboration}),\ }\bibfield  {title} {\bibinfo
  {title} {{The Astropy Project}: {B}uilding an open-science project and status
  of the v2.0 core package},\ }\href {https://doi.org/10.3847/1538-3881/aabc4f}
  {\bibfield  {journal} {\bibinfo  {journal} {Astron. J.}\ }\textbf {\bibinfo
  {volume} {156}},\ \bibinfo {pages} {123} (\bibinfo {year} {2018})},\ \Eprint
  {https://arxiv.org/abs/1801.02634} {arXiv:1801.02634}\BibitemShut {NoStop}%
\bibitem [{\citenamefont {Foreman-Mackey}(2016)}]{corner}%
  \BibitemOpen
  \bibfield  {author} {\bibinfo {author} {\bibfnamefont {D.}~\bibnamefont
  {Foreman-Mackey}},\ }\bibfield  {title} {\bibinfo {title} {{corner.py:
  Scatterplot matrices in Python}},\ }\href
  {https://doi.org/10.21105/joss.00024} {\bibfield  {journal} {\bibinfo
  {journal} {J. Open Source Software}\ }\textbf {\bibinfo {volume} {1}},\
  \bibinfo {pages} {24} (\bibinfo {year} {2016})}\BibitemShut {NoStop}%
\bibitem [{\citenamefont {Macleod}\ \emph {et~al.}(2021)\citenamefont
  {Macleod}, \citenamefont {Areeda}, \citenamefont {Coughlin}, \citenamefont
  {Massinger},\  and\ \citenamefont {Urban}}]{gwpy}%
  \BibitemOpen
  \bibfield  {author} {\bibinfo {author} {\bibfnamefont {D.~M.}\ \bibnamefont
  {Macleod}}, \bibinfo {author} {\bibfnamefont {J.~S.}\ \bibnamefont {Areeda}},
  \bibinfo {author} {\bibfnamefont {S.~B.}\ \bibnamefont {Coughlin}}, \bibinfo
  {author} {\bibfnamefont {T.~J.}\ \bibnamefont {Massinger}},  and\ \bibinfo
  {author} {\bibfnamefont {A.~L.}\ \bibnamefont {Urban}},\ }\bibfield  {title}
  {\bibinfo {title} {{GWpy: A Python package for gravitational-wave
  astrophysics}},\ }\href
  {https://doi.org/https://doi.org/10.1016/j.softx.2021.100657} {\bibfield
  {journal} {\bibinfo  {journal} {SoftwareX}\ }\textbf {\bibinfo {volume}
  {13}},\ \bibinfo {pages} {100657} (\bibinfo {year} {2021})}\BibitemShut
  {NoStop}%
\bibitem [{\citenamefont {{LIGO Scientific Collaboration}}(2018)}]{lalsuite}%
  \BibitemOpen
  \bibfield  {author} {\bibinfo {author} {\bibnamefont {{LIGO Scientific
  Collaboration}}},\ }\href {https://doi.org/10.7935/GT1W-FZ16} {\bibinfo
  {title} {{LIGO} {A}lgorithm {L}ibrary---{LALS}uite}},\ \bibinfo
  {howpublished} {free software (GPL)} (\bibinfo {year} {2018})\BibitemShut
  {NoStop}%
\bibitem [{\citenamefont {Hunter}(2007)}]{matplotlib}%
  \BibitemOpen
  \bibfield  {author} {\bibinfo {author} {\bibfnamefont {J.~D.}\ \bibnamefont
  {Hunter}},\ }\bibfield  {title} {\bibinfo {title} {{Matplotlib: A {2D}
  graphics environment}},\ }\href {https://doi.org/10.1109/MCSE.2007.55}
  {\bibfield  {journal} {\bibinfo  {journal} {Comput. Sci. Eng.}\ }\textbf
  {\bibinfo {volume} {9}},\ \bibinfo {pages} {90} (\bibinfo {year}
  {2007})}\BibitemShut {NoStop}%
\bibitem [{\citenamefont {Virtanen}\ \emph {et~al.}(2020)\citenamefont
  {Virtanen}, \citenamefont {Gommers}, \citenamefont {Oliphant}, \citenamefont
  {Haberland}, \citenamefont {Reddy}, \citenamefont {Cournapeau}, \citenamefont
  {Burovski}, \citenamefont {Peterson}, \citenamefont {Weckesser},
  \citenamefont {Bright}, \citenamefont {{van der Walt}}, \citenamefont
  {Brett}, \citenamefont {Wilson}, \citenamefont {Millman}, \citenamefont
  {Mayorov} \emph {et~al.}}]{scipy}%
  \BibitemOpen
  \bibfield  {author} {\bibinfo {author} {\bibfnamefont {P.}~\bibnamefont
  {Virtanen}}, \bibinfo {author} {\bibfnamefont {R.}~\bibnamefont {Gommers}},
  \bibinfo {author} {\bibfnamefont {T.~E.}\ \bibnamefont {Oliphant}}, \bibinfo
  {author} {\bibfnamefont {M.}~\bibnamefont {Haberland}}, \bibinfo {author}
  {\bibfnamefont {T.}~\bibnamefont {Reddy}}, \bibinfo {author} {\bibfnamefont
  {D.}~\bibnamefont {Cournapeau}}, \bibinfo {author} {\bibfnamefont
  {E.}~\bibnamefont {Burovski}}, \bibinfo {author} {\bibfnamefont
  {P.}~\bibnamefont {Peterson}}, \bibinfo {author} {\bibfnamefont
  {W.}~\bibnamefont {Weckesser}}, \bibinfo {author} {\bibfnamefont
  {J.}~\bibnamefont {Bright}}, \bibinfo {author} {\bibfnamefont {S.~J.}\
  \bibnamefont {{van der Walt}}}, \bibinfo {author} {\bibfnamefont
  {M.}~\bibnamefont {Brett}}, \bibinfo {author} {\bibfnamefont
  {J.}~\bibnamefont {Wilson}}, \bibinfo {author} {\bibfnamefont {K.~J.}\
  \bibnamefont {Millman}}, \bibinfo {author} {\bibfnamefont {N.}~\bibnamefont
  {Mayorov}},  \emph {et~al.},\ }\bibfield  {title} {\bibinfo {title} {{SciPy}
  1.0: {F}undamental algorithms for scientific computing in {Python}},\ }\href
  {https://doi.org/10.1038/s41592-019-0686-2} {\bibfield  {journal} {\bibinfo
  {journal} {Nat. Methods}\ }\textbf {\bibinfo {volume} {17}},\ \bibinfo
  {pages} {261} (\bibinfo {year} {2020})}\BibitemShut {NoStop}%
\bibitem [{dcc(2020)}]{dccGW190814}%
  \BibitemOpen
  \href@noop {} {\bibinfo {title} {{LIGO Document Control Center, GW190814
  parameter estimation samples}}},\ \bibinfo {howpublished}
  {\url{https://dcc.ligo.org/LIGO-P2000183/public}} (\bibinfo {year}
  {2020})\BibitemShut {NoStop}%
\end{thebibliography}%
